\documentstyle[12pt,aasms4]{article}

\lefthead{Bekki Kenji \& Yasuhiro Shioya}
\righthead{Stellar populations in gas-rich galaxy mergers}

\begin{document}
\title{Stellar populations in gas-rich galaxy mergers  I.
  Dependence on star formation history}

\author{Kenji Bekki \& Yasuhiro Shioya\altaffilmark{1}} 
\affil{Astronomical Institute, Tohoku University, Sendai, 980-77, Japan} 

\altaffiltext{1}{Center for Interdisciplinary Research, Tohoku University, 
Sendai, 980-77, Japan}

\begin{abstract}

We  investigate  the nature of stellar populations of major galaxy mergers 
between late-type spirals considerably abundant in interstellar medium
by performing numerical simulations designed 
to solve both the dynamical and chemical evolution in a self-consistent 
manner.
We particularly consider that 
the star formation history of galaxy mergers is a crucial 
determinant  for the nature of stellar populations of merger remnants, 
and therefore
investigate how the difference in  star formation history between galaxy mergers  
affects the chemical evolution of galaxy mergers.
We found that the rapidity of star formation, which is defined
as the ratio of the dynamical time-scale to the time-scale of gas consumption
by star formation,
is the most important determinant for a number of fundamental characteristics of
stellar populations of  merger remnants.
Main results obtained in this study are the following five.
(1) A galaxy merger with more rapid star formation  becomes elliptical with
larger mean metallicity.
This is  primarily because in
the merger with more rapid star formation, a smaller amount of metal-enriched 
gas is tidally stripped away during merging and consequently 
a larger amount of the gas
can be converted to stellar component.  
This result demonstrates that the origin of the color-magnitude relation
of elliptical galaxies can be  closely associated with the details of
merging dynamics which depends on the rapidity of star formation 
in  galaxy mergers.
(2) Negative metallicity gradient fitted reasonably well by power-low can 
be reproduced by  dissipative galaxy mergers with star formation.
The magnitude of metallicity gradient is larger for an elliptical galaxy
formed by galaxy merging with less rapid star formation. 
(3) Absolute magnitude of metallicity gradient  correlates  with that of 
age gradient in  galaxy mergers
in the sence that a merger remnant with steeper negative metallicity gradient 
is more likely to show
steeper age gradient.
(4) The outer part of stellar populations is both older and less  metal-enriched
than nuclei in an elliptical galaxy formed by galaxy merging with
less rapid star formation. 
Moreover, the metallicity of the  outer part of gaseous component for some models
with less rapid star formation is appreciably smaller than that of stellar one. 
This result implies that the origin of metal-poor hot gaseous $X$-ray halo in
real elliptical galaxies can be essentially  ascribed to the dynamics
of dissipative
galaxy merging.
(5) Irrespectively of the rapidity of star formation,
the epoch of galaxy merging affects both the mean stellar metallicity 
and mean stellar age of merger remnants:
Later galaxy mergers are more likely to
become ellipticals with both younger and more metal-enriched
stellar populations.
This result reflects the fact that in the later mergers, 
a larger  amount of
more metal-enriched
interstellar gas is preferentially converted into younger stars in the
later star formation  triggered by galaxy merging. 
These five results clearly demonstrate that even the chemical  
evolution  of elliptical galaxies can be strongly affected by  the details
of dynamical evolution of galaxy merging, which is furthermore 
determined  by the rapidity of star formation of galaxy mergers.
In particular, tidal stripping of interstellar gas and total amount of
gaseous dissipation during galaxy merging
are  demonstrated to play a vital role in determining a number of
chemical properties
of merger remnants.
Based upon these results, we adopt a specific assumption of
the luminosity dependence of the rapidity of star formation
and thereby  discuss how successfully
the present merger model can reproduce a number of 
fundamental chemical, photometric, and spectroscopic 
characteristics  of elliptical galaxies.

\end{abstract}


\section{Introduction}

Elliptical galaxies have been generally considered
to be old, coeval and homogeneous systems passively
evolving after the single initial burst of star formation associated
with dissipative galaxy formation.
This classical picture of elliptical galaxy formation
appears to have been supported by
the considerably tight color-magnitude relation
of elliptical galaxies 
( Bower, Lucey, \& Ellis 1992;
Ellis et al. 1997)
and by relatively smaller  redshift evolution  of photometric properties
of elliptical galaxies
(Arag$\rm \acute{o}$n-Salamanca et al. 1993; Franx \& van Dokkum 1996).
A growing number of recent observational results, however, shed a strong
doubt on this long-standing  
view of elliptical galaxy formation,
and suggest that there is  great variety of star formation history
between elliptical galaxies,
such as the epoch of major star formation, the duration and efficiency
of star formation
(Worthey, Faber, \&
Gonzalez 1992; Matteuchi 1994; Faber et al. 1995; Bender 1996;
Worthey, Trager, \& Faber 1996).
This tendency that elliptical galaxies show diversity in star formation
history and nevertheless can actually keep the tightness of the color-magnitude
relation is considered to be quite mysterious
and thus to provide any theoretical models with a valuable insight on the elliptical
galaxy formation.
Such kind of mysterious nature observed in elliptical
galaxies is demonstrated to hold equally 
for the dynamical  and kinematical properties  
of elliptical galaxies. For example,
considerably small thickness of the fundamental plane of elliptical galaxies
implies a rather smaller range of admitted dynamical state  
of the galaxies (Djorgovski \& Davis 1987; Dressler et al. 1987 ;
Djorgovski, Pahre, \& de Carvalho 1996)
 whereas the morphological 
dichotomy between boxy-disky elliptical galaxies
(Kormendy \& Bender 1996) and the projected density profile systematically
 departing  from de Vaucouleurs $R^{1/4}$ law 
(Caon, Capaccioli, \& D'Onofrio 1993) 
show  a great variety of major orbit families consisting the galaxies.
These fundamental characteristics that elliptical galaxies show both
diversity and uniformity in their chemical, photometric and dynamical properties
have imposed  some stringent  but valuable constraints on any theoretical models of
elliptical galaxy formation.
What is the most vital in challenging the origin of elliptical galaxy formation
in this kind of situation
is to investigate whether or not both the chemical and photometric 
properties and dynamical and kinematical ones can be reproduced successfully by
a specific model of galaxy formation in a reasonably self-consistent manner.
The previous theoretical models
 addressing this important issue on elliptical galaxy formation
are divided basically into two categories: The dissipative galactic collapse 
model (e.g., Larson 1976; Carlberg 1984) 
and the galaxy merger model (e.g., Toomre \& Toomre 1972).
As is suggested by Kormendy \& Sanders (1992),
these two dominant and apparently competing scenarios for elliptical galaxy
formation are now converging, thus it would be
crucial  to construct  one more realistic and sophisticated model
of elliptical galaxy formation.
Although there are a large number of important studies exploring the origin of
elliptical galaxy formation along the dissipative collapse scenario,
especially in the context of the nature of stellar populations
(e.g., Arimoto \& Yoshii 1987),
we here restrict ourselves to the merger scenario  of elliptical galaxy 
formation.

 Recent extensive studies of merger models of elliptical galaxy formation,
mostly based upon numerical simulations, $appear$ to have succeeded in 
resolving most of the outstanding problems related to dynamical and kinematical
properties of elliptical galaxies, such as the phase space density (Ostriker 1980;
Carlberg 1986) and
kinematical misalignment 
(Franx, Illingworth, \& de Zeeuw 1991; Barnes 1992), 
by invoking the inclusion of bulge component, gaseous dissipation, and multiplicity
of galaxy merging (Hernquist, Spergel, \& Heyl 1993; Weil \& Hernquist 1996;
Barnes \& Hernquist 1996).
Although it would be safe to say that the galaxy merging
between two late-type spirals  is one of the most promising
candidates explaining more clearly the origin of elliptical galaxies 
at least in the context of the $dynamical$ $and$ $kinematical$ $properties$,
however,
there still remain  a number of unresolved  and apparently serious problems concerning 
the merger model (e.g., van den Bergh 1995).
One of the most crucial
problems among these is on whether  the fundamental characteristics of
stellar populations of elliptical galaxies can be reproduced reasonably well
by galaxy merging between two late-types spirals.
Surprisingly, there are only a few works addressing this critical issue for 
the merger model, probably because 
it is considered to be rather difficult
to solve the chemical evolution of galaxy mergers
in which a number of competing physical processes are expected
to affect strongly the chemical evolution of galaxy mergers. 
White (1980) and Mihos \& Hernquist (1994) found  that 
the stellar populations of progenitor disks 
are not  mixed so well even by the violent relaxation during galaxy merging
and consequently the metallicity gradient
of progenitor disks is not so drastically washed out. 
The metallicity gradient of merger remnant is furthermore found not 
to  be fitted by 
power law  observed in elliptical galaxies (Mihos \& Hernquist 1994). 
Schweizer \& Seitzer (1992) discussed whether or not the bluer integrated $UBV$ color
of elliptical galaxies with morphologically fine structure can be explained by
secondary starburst induced by  major disk-disk galaxy mergers.
Kauffmann \& Charlot (1997) construct  a semi-analytic model of 
elliptical galaxy formation, which is based upon
the hierarchical clustering in CDM universe and includes  rather simple
 chemical enrichment  process, and thereby  
 demonstrate that the origin of the color-magnitude relation of elliptical
galaxies can be reproduced successfully even in the CDM model of galaxy formation
(See also Baugh, Cole \& Frenk 1996.).
Thus, since  there are only a few works addressing chemical and photometric
properties  for 
the merger model, it is essential for the merger model to investigate more throughly 
the fundamental chemical and photometric properties of merger remnants,
including the origin of color-magnitude relation 
(Faber 1973; Visvanathan \& Sandage 1977), age and metallicity gradient (Peletier 
et al. 1990; Davies et al 1991),
$\rm Mg_{2} - \sigma$ relation (Burstein et al. 1988),
age-metal-conspiracy in stellar populations (Faber et al. 1995;
Worthey et al. 1996), 
luminosity dependence of the line ratio
 [Mg/Fe] (Worthey et al. 1992),  metal-poor gaseous $X$-ray  
halo (Matsumoto et al. 1997), 
and the substantially metal-enriched galactic nuclei at higher redshift
(Hamann \& Ferland 1993).

  What should be recognized foremost in investigating the nature of stellar populations
in  merger remnants is that a glowing number of observational results have been
accumulated which suggest the relatively earlier formation of elliptical galaxies.
Tightness of the color-magnitude relation in the  cluster of galaxies (Bower et al. 1992,
Ellis et al 1996),
relatively smaller photometric evolution of cluster ellipticals (Arag$\rm 
 \acute{o}$n-Salamanca
et al. 1993), and the 
redshift evolution
of the fundamental plane (Franx \& van Dokkum 1996) 
all suggest the $typical$ formation epoch of elliptical
galaxies is  earlier  than 2 in redshift.
Furthermore, as is suggested by Kormendy \& Sanders (1992),
the fact that no galaxy in the $K$-band
survey of Cowie et al. (1994) shows the global color resembling that of
the Arp 220, which is considered
to be  ongoing mergers and forming ellipticals, implies that
the formation epoch of elliptical galaxies should be earlier than 1.0 in redshift. 
Silva \& Bothun (1997) revealed that 
the fraction of mass of stellar populations with intermediate age to total mass in 
elliptical galaxies with morphologically fine structure is less than 15 percent.
These results imply that if  elliptical galaxies are formed by galaxy
merging, the epoch of galaxy merging should be relatively
earlier and furthermore 
that the precursor disks of galaxy mergers may be extremely abundant in
interstellar medium compared with the present spirals.
Recent high quality imaging using $Hubble$ $Space$ $Telescope$
($HST$)  has revealed that a  larger number of 
galaxies at faint magnitude are interacting/merging galaxies 
(e.g., van den Bergh et al. 1996), indicating furthermore that the potential
candidate for elliptical
galaxies formed by galaxy merging  are ubiquitous in higher redshift universe.
Hence it is quite  reasonable and essential to study the nature of stellar populations
of higher redshift  galaxy mergers between  disk galaxies with the gas mass fraction
larger than 0.2,
which is a typical value of the present late-type spirals,
and thereby  to confirm whether or not  
 elliptical galaxies can be  formed $actually$  by galaxy merging.

The purpose of this paper is to explore the nature of the stellar populations
of a gas-rich disk merger which is considered to  be occurred the most frequently
in the high redshift universe. 
We particularly investigate how successfully galaxy mergers between gas-rich
spirals
can reproduce a number of fundamental chemical, photometric,
and spectroscopic  properties
of elliptical galaxies.
The layout of this paper is as follows.
In \S 2, we summarize  numerical models used in the
present study and describe in detail   methods for
analyzing the stellar populations produced by dissipative galaxy
mergers with star formation.
In \S 3, we demonstrate how  a number of fundamental  characteristics 
of stellar populations in merger remnants are  
affected by  the star formation history of 
dissipative galaxy merging.
In \S 4, we discuss how successfully
the present merger model can reproduce a number of
observational results concerning the chemical, photometric, and
spectroscopic properties of elliptical galaxies.
In this section, we also point out the advantages and disadvantages of
galaxy mergers in explaining both the chemical, photometric, 
and spectroscopic properties
and dynamical and kinematical ones in real elliptical galaxies.  
The conclusions of the present study are given in \S 5.

\section{Model}
Dynamical evolution of dissipative galaxy mergers with
star formation is generally considered to be 
highly complex principally because only a smaller amount of interstellar gas can 
drastically change the degree of violent relaxation, the details of
redistribution of angular
momentum in gaseous and stellar component, and the total amount of mass 
transfered to the central region of merger remnants (Barnes \& Hernquist 1992, 1996).
Basically the transfer and mixing of heavy elements ejected from stellar component 
are controlled  by the above dynamical processes of galaxy merging in a considerably
complicated way, 
we accordingly could  not be allowed simply to invoke the Simple one-zone model
in analyzing the chemical and photometric evolution of merging galaxies.  
Thus, in order to analyze the nature of stellar populations
produced in such a complex situation of  dissipative galaxy merging,
we must solve both dynamical and chemical evolution in a admittedly self-consistent
manner.
In the present study,
dynamical evolution of collisional component (interstellar gas) and  
collisionless  one (dark halo and stars) is  solved by a specific
N-body method and then a number of  characteristics of stellar populations 
at given points
in the merger remnants are 
calculated  based on the derived information about 
the position, velocity, age, and metallicity of each stellar particle.
Firstly we describe the numerical model 
including the initial conditions of the mergers,
the prescriptions of dissipative process, and the model for star formation
in \S 2.1. 
Secondly we give the method for analyzing the chemical enrichment process
during mergers
and the photometric properties of the remnants in \S 2.2.
Thirdly, we describe the main points of analysis of the present study in \S 2.3. 
Lastly, we give values of each parameter in each model
in \S 2.4. 

\subsection{Numerical model}

\subsubsection{Initial conditions}

 We construct  models of galaxy mergers between gas-rich 
 disk galaxies with equal mass by using Fall-Efstathiou model (1980).
 The total mass and the size of a progenitor disk are $M_{\rm d}$
 and $R_{\rm d}$, respectively. 
 From now on, all the mass and length are measured in units of
  $M_{\rm d}$ and  $R_{\rm d}$, respectively, unless specified. 
  Velocity and time are 
  measured in units of $v$ = $ (GM_{\rm d}/R_{\rm d})^{1/2}$ and
  $t_{\rm dyn}$ = $(R_{\rm d}^{3}/GM_{\rm d})^{1/2}$, respectively,
  where $G$ is the gravitational constant and assumed to be 1.0
  in the present study. 
  If we adopt $M_{\rm d}$ = 6.0 $\times$ $10^{10}$ $ \rm M_{\odot}$ and
  $R_{\rm d}$ = 17.5 kpc as a fiducial value, then $v$ = 1.21 $\times$
  $10^{2}$ km/s  and  $t_{\rm dyn}$ = 1.41 $\times$ $10^{8}$ yr,
  respectively.
  In the present model, the rotation curve becomes nearly flat
  at  0.35  radius with the maximum rotational velocity $v_{\rm m}$ = 1.8 in
  our units.
  The corresponding total mass $M_{\rm t}$ and halo mass $M_{\rm h}$
  are 3.8 and 2.8 in our units, respectively.
  The radial ($R$) and vertical ($Z$) density profile 
  of a  disk are  assumed to be
  proportional to $\exp (R/R_{0}) $ with scale length $R_{0}$ = 0.2
  and to  ${\rm sech}^2 (Z/Z_{0})$ with scale length $Z_{0}$ = 0.04
  in our units,
  respectively.
  The mass density  of halo component is truncated at  1.2 in our units
  and its velocity dispersion at a given point
  is set to be isotropic and given
  according to the  virial theorem.
  In addition to the rotational velocity made by the gravitational
  field of disk and halo component, the initial radial and azimuthal velocity
  dispersion are given to disk component according
  to the epicyclic theory with Toomre's parameter (\cite{bt87}) $Q$ = 1.0.
  This adopted value for $Q$ parameter
is appreciably smaller compared with the value required 
 for stabilizing the initial disk
against the non-axisymmetric dynamical instability (e.g. bar instability).
The reason for this adoption is that the initial disk is assumed to
be composed mostly of interstellar gas and thus random kinetic energy
in the disk is considered to be rather smaller because of gaseous
dissipation in  the disk. 
  The vertical velocity dispersion at given radius 
  are set to be 0.5 times as large as
  the radial velocity dispersion at that point, 
  as is consistent  with 
  the observed trend  of the Milky Way (e.g., Wielen 1977).
 As is described above, the present initial disk model does not
 include any remarkable bulge component, and accordingly corresponds to
 `purely'  late-type spiral without galactic bulge. Although it is
 highly possible that  galactic bulges greatly affect the chemical evolution
 of galaxy mergers, we however investigate
 this issue in our future papers.
 It could be reasonable to consider 
 that the present initial disk with a considerably larger amount of
 interstellar gas and without pronounced galactic bulges
 is one of  candidates of  precursors of typical
 elliptical galaxies which are
 considered to be formed at relatively higher redshift ($z$ $>$ 2).

  The collisional and dissipative nature 
  of the interstellar medium is  modeled by the sticky particle method
  (\cite{sch81}).
It should be emphasized here that this discrete cloud model can at best represent
the $real$ interstellar medium of galaxies  in a schematic way. 
As is modeled by McKee \& Ostriker (1977),
the interstellar medium can be considered to be  
 composed mainly of `hot', `warm', and `cool'
gas,
each of which mutually
interacts hydrodynamically 
 in a rather  complicated way.
 Actually, these considerably complicated nature of
interstellar medium in  disk galaxies would not be
  so simply modeled by the `sticky
particle' method in which gaseous dissipation is modeled by ad hoc
cloud-cloud collision: Any existing numerical method probably could
not model the $real$ interstellar medium in an admittedly proper
way. 
In the present study, as a compromise,
we only try to address some important aspects of hydrodynamical
interaction between interstellar medium in disk galaxies and in
dissipative mergers. 
More elaborated numerical modeling for real interstellar medium
would be  necessary for 
our further understanding of dynamical evolution 
in dissipative galaxy mergers. 
  To mimic the
  galaxy mergers which are 
occurred at   higher redshift and thus very dissipative
because of a  considerably larger amount of  interstellar gas in the 
progenitor disks,
  we assume that the fraction of gas mass in
  a disk is set to be 1.0 initially.
Actually, the gas mass fraction in precursor disks of a merger
is  different between galaxy mergers  and depends on the epoch of
the merging.
Although this difference probably could yield a great variety of chemical
and dynamical structures in merger remnants, we do not intend to
consider this important difference for simplicity in the present  paper
and will address in our future paper. 
  The size  of the clouds is set to be 3.5 $\times$ $10^{-3}$ in our units 
  in the present simulations. 
  The radial and tangential restitution coefficient for cloud-cloud
  collisions are
  set to be 0.5 and
  0.0, respectively.
  The  number of particles of halo and the gaseous component
  for an above  isolated galaxy are
  5000 and 10000, respectively.

    Numerical simulations of galaxy mergers are divided into two categories in
the present study: Pair mergers between  two disks
with a parabolic encounter, and multiple mergers between  five disks. 
    In all of the simulations of pair mergers, the orbit of the two disks is set to be
    initially in the $xy$ plane and the distance between
    the center of mass of the two disks,
represented by $r_{\rm in}$,
  is  assumed  to be the  free parameter 
which controls the epoch of galaxy merging.
    The pericenter
    distance, represented by $r_{\rm p}$, is also
assumed to be the  free parameter which controls the initial
total orbital  angular momentum of galaxy mergers.
The eccentricity
is  set to be 1.0 for all models of pair mergers,
meaning that the encounter of galaxy merging is parabolic.
    The spin of each galaxy in a  pair merger
is specified by two angle $\theta_{i}$ and
    $\phi_{i}$, where suffix  $i$ is used to identify each galaxy.
    $\theta_{i}$ is the angle between the $z$ axis and the vector of
    the angular momentum of a disk.
    $\phi_{i}$ is the azimuthal angle measured from $x$ axis to
     the projection of the angular momentum vector of a disk on
    to $xy$ plane. 
The value of each parameter,  $\theta_{i}$, $\phi_{i}$, $r_{\rm p}$ 
and $r_{\rm in}$ for each model is described later. 
In the simulations of multiple mergers, the initial position of
each progenitor disk is set to be distributed 
  randomly within a sphere with radius 6.0 in
our units, and the initial velocity dispersion of each disk (that is,
the random motion of each galaxy in the sphere) is set to be
distributed  in such a
way that the ratio of the total kinematical 
energy to the total potential energy in the system is  0.25.  
The time when the progenitor disks merge completely and reach  the
dynamical equilibrium is less than 15.0 in our units for most of
models and does not depend so
strongly on the  history of star formation in the  present calculations.

\subsubsection{Global star formation}

    Star formation
     is modeled by converting  the collisional
    gas particles
    into  collisionless new stellar particles according to the algorithm
    of star formation  described below.
    We adopt the Schmidt law (Schmidt 1959)
    with exponent $\gamma$ = 2.0 (1.0  $ < $  $\gamma$
      $ < $ 2.0, \cite{ken89}) as the controlling
    parameter of the rate of star formation.
    The amount of gas 
    consumed by star formation for each gas particle
    in each time step, 
    $\dot{M_{\rm g}}$, 
is given as:
    \begin{equation}
      \dot{M_{\rm g}} \propto  C_{\rm SF} \times 
 {(\rho_{\rm g}/{\rho_{0}})}^{\gamma - 1.0}
    \end{equation}
    where $\rho_{\rm g}$ and $\rho_{0}$
    are the gas density around each gas particle and
    the mean gas density at 0.48 radius  of 
    an initial disk, respectively.
    This star formation model is similar to that of Mihos, Richstone, \& Bothun  (1992).
    In order to avoid a large number of new stellar particles with
different mass, we convert one gas particle into one stellar one
according to the following procedure.
First we give each gas particle the probability, $P_{\rm sf}$,
 that the gas particle
is converted into stellar one, by setting the $P_{\rm sf}$ to be proportional
to the  $\dot{M_{\rm g}}$  in equation (1) estimated for the gas particle. 
Then we draw the random number to determine whether or not the gas particle
is totally converted  into one new star. 
This method of star formation enables us to control the rapidity of star formation
without increase of particle number in each simulation thus to maintain
the numerical accuracy in  each simulation. 
    The $C_{\rm SF}$ in the equation (1)
is the parameter that controls the rapidity of 
    gas consumption by star formation:
    The larger the $C_{\rm SF}$ is, the more rapidly the gas particles 
    are converted to new stellar particles.
As a result of this, total amount of gaseous dissipation is 
also controlled by this parameter $C_{\rm SF}$:
The more amount of kinetic energy of gas particles is dissipated away by
cloud-cloud collision for the models with smaller $C_{\rm SF}$.
This parameter $C_{\rm SF}$ is meant to be  proportional to
the ratio of dynamical time-scale of the system to the time-scale of
gas consumption  by star formation.
Furthermore the equation (1) states that
a larger number of stellar particles are created at the
regions  where the local gas density become larger owing to the onset
of local Jeans instability.
The positions and velocity of the new stellar particles are set to 
be the same as those of original gas particles.
As is described above, in the present study, we do not explicitly
include the `threshold density' of star formation,
which is demonstrated to be associated with the growth
of local gravitational
instability with  relatively smaller wavelength, that is, the Toomre's 
$Q$ parameter (Kennicutt 1989), for isolated disk galaxies.
This is because
the threshold criterion, which is derived only for  calmly evolving
galactic disks,
would not so simply be applied to the present merger model,
 in which 
the time-scale that the disk can evolve without strong dynamical 
perturbation is relatively smaller (less than  10 dynamical time for typical models).

 In  the present study, we do not intend to include any `feedback
effects' of star formation such as 
thermal and dynamical heating of interstellar medium driven by
supernovae, firstly because such  inclusion could
prevent us from deducing more clearly the important roles of the rapidity
of star formation and secondly because there still remains a great
uncertainty concerning the numerical  implementation of the `feedback
effects' (Katz 1992; Navarro \& White 1993). 
Accordingly, as is described above, the `star formation' in this preliminary
study only means
the formation of collisionless particles and does not literally mean
the actual and realistic series of star formation.
This kind of
modeling for star formation is rather oversimplified so that
we can only address some important aspects of the roles of
star formation in generating chemical and dynamical characteristics
of merger remnants.  
However, we believe that since the main points of the present study
are only the relatively global and average characteristics of  chemical and
photometric properties,
even the rather simple model adopted in this study makes it possible to
grasp some essential ingredients of the roles of  `star formation'
in producing a number of important characteristics of
 stellar populations  in merging galaxies.
More extensive studies on this subject will be done in our future papers
by using more elaborated model for star formation.

   All the calculations related to 
the above dynamical evolution  including the dissipative
dynamics, star formation, and gravitational interaction between collisionless
and collisional component 
 have been carried out on the GRAPE board
   (\cite{sug90})
   at Astronomical Institute of Tohoku University.
   The parameter of gravitational softening is set to be fixed at 0.04  
   in all the simulations. The time integration of the equation of motion
   is performed by using 2-order
   leap-flog method. Energy and angular momentum  are conserved
within 1 percent accuracy in a test collisionless merger simulation.
Most of the  calculations are set to be stopped at T = 25.0 in our units
unless specified.

\subsection{Method for analysis of stellar population}

Chemical and photometric properties such as global colors and metallicity
gradient in elliptical galaxies
depend critically on how the metallicity and age of  stellar component
are distributed in the galaxies.
Accordingly, we first analyze the distribution of stellar age and 
that of stellar metallicity in merger remnant for each model in
order to grasp the characteristics of stellar population 
in merger   remnants. 
In the present study, 
these age and metallicity distribution are calculated 
based on the age and metallicity assigned to  
each stellar particle, as described in detail later.
The outline for this calculation is as follows. 
First we derive the  distribution of  stellar age and that of stellar 
metallicity 
by assigning  age and metallicity to stellar particles according as the
law of chemical enrichment applied to this study.
Next, by using the stellar  population synthesis method,
we calculate  the photometric and spectroscopic properties 
based on the derived distribution
of age and metallicity in merger remnant.

\subsubsection{Chemical enrichment}

 Chemical enrichment through star formation during galaxy merging
is assumed to proceed both locally and instantaneously in the present study.
The model for analyzing
metal enrichment of each gas  and stellar particle 
is as follows.
First, as soon as a gas particle is converted into a new stellar one by
star formation, we search neighbor gas particles locating within
$R_{\rm che}$ from the position of the new stellar particle
 and then  count
the number of the neighbor gas particles, $N_{\rm gas}$.
This  $R_{\rm che}$ is referred to as 
chemical mixing length in the present paper,
and represents the region within which the neighbor
gas particles are polluted by metals ejected from the new stellar particle.
The value of  $R_{\rm che}$  relative to the typical size of
a galaxy could be different between galaxies, accordingly the
value of  $R_{\rm che}$  is considered to be free parameter in
the present study.
The values of  $R_{\rm che}$ examined the most extensively 
in the present study are 
0.4, which is slightly smaller than the effective radius of  typical
merger remnants in the present study, 0.1, and 0.02, which is the half
of the gravitational softening length.
Next we assign the metallicity of original
gas particle to  the new stellar particle and increase 
the metals of the each neighbor gas particle according to the following 
equation about the chemical enrichment:
  \begin{equation}
  \Delta M_{\rm Z} = \{ Z_{i}R_{\rm met}m_{\rm s}+(1.0-R_{\rm met})
 (1.0-Z_{i})m_{\rm s}y_{\rm met} \}/N_{\rm gas} 
  \end{equation}
where the $\Delta M_{\rm Z}$ represents the increase of metal for each
gas particle. $ Z_{i}$, $R_{\rm met}$, $m_{\rm s}$,
and $y_{\rm met}$  in the above equation represent
the metallicity of the new stellar particle (or that of original gas particle),
the fraction of gas returned to interstellar medium,  the
mass of the new star, and the chemical yield, respectively.
The value of the $R_{\rm met}$ and that of $y_{\rm met}$ are set to
be 0.3 and 0.03, respectively.
Furthermore, the time, $t_{i}$, 
when the new stellar particle is created, is 
 assigned to the new stellar particle in order to calculate the 
photometric evolution of merger remnants, as is described later.
To verify the accuracy of the above  treatment 
(including numerical code) for 
chemical enrichment process,
we checked whether or not the following conservation law of chemical 
enrichment is satisfied for each time step in each test simulation:
  \begin{equation}
 \sum_{\rm star} m_{\rm s}Z_{i} + \sum_{\rm gas}
  m_{\rm g}Z_{i} = y_{\rm met}
 \sum_{\rm star} m_{\rm s}
  \end{equation}
where  $ m_{\rm g}$,  $ m_{\rm s}$, and $ Z_{i}$
 are  the mass of each gas particle,
that of each stellar one, and the metallicity
of each particle,  respectively, and the  summation ($\sum$) is done
for all the gas particles or stellar ones. 
Strictly speaking, the above equation  holds when the value of
$R_{\rm met}$ is 0.0.
Thus, in testing the validity of the present code of chemical enrichment,
we set the  value of  $R_{\rm met}$ to be 0.0 and then perform a simulation
for the test.
We confirmed that the above equation is nearly exactly satisfied
in our test simulations and furthermore that even if the  $R_{\rm met}$ is
not  0.0, the difference in the value of
total metallicity between the left and right side in the above equation
is negligibly small.

\subsubsection{Population synthesis}

It is assumed in the present study
that the spectral energy distribution (SED) of a model galaxy is 
a sum of the SED of  stellar particles. 
The SED of each  stellar particle is assumed to be  
a simple stellar population (SSP) that is  
a coeval and chemically homogeneous assembly of stars. 
Thus the monochromatic flux of a galaxy with  age $T$,
$F_{\lambda}(T)$,  is described as 
\begin{equation}
F_{\lambda}(T) = \sum_{\rm star} F_{{\rm SSP},\lambda}(Z_{i},
{\tau}_{i}) \times m_{\rm s} \; ,
\end{equation}
where $F_{{\rm SSP},\lambda}(Z_{i},{\tau}_{i})$ and $m_{\rm s}$ 
 are  a monochromatic flux of SSP 
of age ${\tau}_{i}$ and metallicity $Z_{i}$, where suffix $i$ identifies 
each stellar particle,  and 
mass of each stellar  particle,  respectively.
The age of SSP, ${\tau}_{i}$, is defined as ${\tau}_{i} = T - t_{i}$, 
where $t_{i}$ is the time when a gas particle convert to a stellar one.
The metallicity of SSP is exactly the same
as that  of the stellar particle, $Z_{i}$, and the summation ($\sum$) in
equation (4) is done  for all 
stellar particles in a model galaxy.

A stellar particle is assumed to be composed of stars whose
age and metallicity are exactly the same as those of the stellar particle
and 
the total mass of the stars is set to be the same as that of
the  stellar particle.
Thus the monochromatic flux of SSP at a given wavelength is defined as
\begin{equation}
F_{{\rm SSP}, \lambda}(Z_{i},{\tau}_{i}) = \int_{M_L}^{M_U} 
\phi (M) f_{\lambda}(M, {\tau}_{i}, Z_{i}) dM \; ,
\end{equation}
where $M$ is mass of a star, $f_{\lambda}(M, {\tau}_{i}, Z_{i})$
 is a monochromatic flux 
of a star with mass $M$, metallicity $Z_{i}$ and age ${\tau}_{i}$.
$\phi (M)$ is a initial mass function (IMF) of stars and 
$M_U$, $M_L$ are upper and lower mass limit of IMF, respectively. 

In this paper, we use the $F_{{\rm SSP}, \lambda}(Z_{i}, {\tau}_{i})$ 
calculated by Tantalo et al. (1996).
The characteristics of Tantalo et al. (1996)'s SSP are 
(1) all evolutionary phase, from the main sequence to the white dwarf or 
C-ignition stage, are included and 
(2) metallicity, $Z$, ranges from 0.0004 to 0.1 and helium content, $Y$, 
satisfies the relation, $Y=2.5Z + 0.230$. 
The library of stellar spectra calculated by Kurucz (1992) 
and the following form of  the Salpeter IMF are used in deriving the
above  $F_{{\rm SSP}, \lambda}(Z_{i}, {\tau}_{i})$:
\begin{equation}
\phi(M) = 
\frac{(x-2)M_L^{x-2}}{1 - \left( \frac{M_L}{M_U} \right)^{x-2}} M^{-x} \; ,
\end{equation}
where $x = 2.35$, $M_L = 0.15 M_{\odot}$ and $M_U = 120 M_{\odot}$. 
In order to estimate the $F_{\rm SSP} (Z_{i}, {\tau}_{i})$ for 
stellar particles with arbitrary $Z_{i}$ and ${\tau}_{i}$, 
we assign interpolated values to each stellar particle
by using  a set of data points tabulated  in Table 2 
of Tantalo et al. (1996). 
The mean stellar metallicity $< Z_* >$ used in the present study is 
defined as 
\begin{eqnarray}
<Z_*> & = & 
\frac{\displaystyle \sum_{\rm star} Z_{i}  m_{\rm s}}
{\displaystyle \sum_{\rm star}  m_{\rm s}} \; . 
\end{eqnarray}
Colors at given radius and radial color gradients are given as follows. 
First we divide the region ranging from 0.5 $R_{\em e}$ to 5.0 $R_{\em e}$
of a merger remnant into 20 annulus whose inner and outer boundary 
are  located at the distance $r_{j}$ and $r_{j+1}$
from the center of the remnant, respectively.
Then the monochromatic flux emitted from the $j$th annulus  
is derived by summing the   $F_{\rm SSP} (Z_{i}, {\tau}_{i})$
of all the particles locating at  $r_{j}$ $<$ $R_{i}$  $<$ $r_{j+1}$,
where $R_{i}$ is the distance between the position of each stellar particle
and the center of the remnant. 
Color gradients are calculated  
by using this  monochromatic flux  for  each annulus.


\subsection{Main points of analysis}
 We consider that the $C_{\rm SF}$ is the most important parameter in the present study,
 principally because the $C_{\rm SF}$ could depend on galactic luminosity.
 The importance of the $C_{\rm SF}$ and its possible dependence on galactic luminosity 
 are described in the  Appendix section.
 In what follows,
we mainly investigate how the rapidity of 
gas consumption by star formation
($C_{\rm SF}$) can determine a number of fundamental chemical,
photometric, and spectroscopic properties  
of merger remnants, and thereby observe
how successfully the present merger model can  reproduce the 
fundamental characteristics of real elliptical galaxies.
Since the $C_{\rm SF}$ is considered to depend on galactic
luminosity in the present study,
this investigation corresponds to addressing indirectly 
the origin of the luminosity-dependent characteristics
of elliptical galaxies.
First we investigate 
the dependence of chemical properties on the $C_{\rm SF}$,
which is the most important parameter in the present study,
in \S 3.1. 
What we investigate the most extensively in
this investigation  are on the mean stellar and gaseous metallicity,
radial gradients of metallicity and age, shapes of isochrome contour, 
and the formation
of substantially metal-enriched galactic nuclei,
in merger remnants.
Second  we  investigate the important roles of other four parameters,  
the chemical mixing length, 
orbit configuration  of galaxy merging,
multiplicity of galaxy merging, and
the epoch of galaxy merging,
in determining the fundamental chemical properties of merger remnants (\S 3.2). 
This sort of multi-parameter study is indispensable for the present study,
in which initial physical conditions of galaxy mergers can be variously different.
In this investigation, we confirm whether or not the important roles of the parameter
 $C_{\rm SF}$ derived in \S 3.1 can be  equally seen in the models with
different chemical mixing length, orbit configuration, multiplicity of mergers,
and merging epoch.
We also observe how these four parameters modify the $C_{\rm SF}$
dependence  derived in \S 3.2 and what fundamental roles the four parameters
play in determining the characteristics of stellar populations. 
Third  we investigate 
the photometric and spectroscopic 
properties of merger remnants
and their dependence on the galactic luminosity, 
by adopting a specific relation between the $C_{\rm SF}$ and galactic
luminosity  and by assigning the mass and scale to  each model with $C_{\rm SF}$,
in \S 3.3.

\subsection{Parameter values of each model}
In the present paper, totally  29  models  
are investigated, and the
parameters and brief summary of the results are given in Table 1. 
Each model of galaxy merger is labeled, for example, as Model B1, according to 
the orbit configuration of galaxy merging and multiplicity of the merging.
Most of models (23 models) are assigned to  pair mergers in Model 
B (standard models), C, D, E, and F
and only
two are assigned  to  multiple mergers in Model  G.   
For comparison, the isolated evolution of a progenitor disk of a galaxy merger
is also investigated and labeled as Model A.
The values of  orbit parameters, $\theta_{1}$
$\theta_{2}$,  $\phi_{1}$, and  $\phi_{2}$ are 
set to be 30.0, 120.0,  90.0, and  0.0, respectively, for Model B and F,
0.0, 120.0, 0.0, and 0.0 for Model C, 0.0, 30.0, 0.0, and 0.0 for Model D,
150.0, 150.0, 0.0, and 0.0 for Model E.
Orbit configuration of Model D and that of Model E correspond to
nearly prograde merger and nearly retrograde one, respectively.
The  Model F with $r_{\rm p}$ = 0.5 corresponds to the galaxy merger
with smaller initial orbital angular momentum.  
The difference between the Model A1 (2) and Model A3 (4) is that
in the Model A3 (4), chemical evolution including star formation
is solved but dynamical evolution is $not$.
Accordingly,
structural and kinematical properties of the disk are set to be  
the same between the initial state and final one in the models, Model A3 and A4.

First, second and third column in the Table 1 denote the model number, 
value of the parameter $C_{\rm SF}$,  and that of $R_{\rm che}$,
respectively, for each model.
The values of $r_{\rm in}$ and $r_{\rm p}$
are given in the fourth and fifth column, respectively.
We give the mean gaseous metallicity ($<Z>$), the mean stellar metallicity
($<Z_{\ast}>$), 
the gas mass fraction $f_{\rm g}$ 
in the sixth,  seventh, and eighth  column, respectively.
The values of the mean gaseous ($<Z_{\rm Sim}>$) and stellar 
($<Z_{\ast ,\rm Sim}>$) metallicity predicted from   the Simple one-zone
model are presented in the ninth and tenth column, respectively.  
The mean epoch of star formation ($<T_{\ast}>$),
which is defined as the average of star formation epoch ($t_{ i}$)
for  all  stellar particles in a merger remnant 
at final state ($T=15$ in our units),
is given in the eleventh column. 
 $T_{M_{\ast}/M = 0.6}$ shown in the twelves column  represents the
 time-scale (in our units)
 that 60 percent of initial gas 
 of progenitor disk galaxies has been  converted
 into stellar particles during a simulation.
The thirteenth  column gives the 
magnitude of metallicity gradient which is defined
as $\Delta \log <Z_{\ast}> / \Delta \log R$ ($R$ is the distance
from the center of merger remnant.) and measured for the region
ranging from $0.5 R_{\rm eff}$ to $5.0  R_{\rm eff}$ ($R_{\rm eff}$
is the effective radius of each model.) 
in the $xy$ projection
for each model.


\placefigure{fig-1}
\placefigure{fig-2}

\section{Results}
 In this section, we observe how the star formation history, in particular, the 
rapidity of star formation, in galaxy mergers can affect the fundamental characteristics
of stellar populations in mergers remnants. 
Before describing the results of merger remnants, we begin to
briefly observe the dynamical
evolution of an isolated disk galaxy with star formation, 
which is considered to be a progenitor of a
galaxy merger in the present study,
 and the dependence of star formation history on the parameter 
$C_{\rm SF}$ in galaxy mergers.
Figure 1 gives  snapshots of an isolated disk model with $C_{\rm SF}$ = 0.35 
(Model A1) at $T$ = 15.0,
in which about 60 percent of initial gas mass is converted into stellar 
component within 6.5 dynamical time. 
In the present isolated disk model,
a stellar bar  sufficiently
develops in the central region of the disk
only after more than 60 percent of initial gas mass has been consumed
by star formation in the disk.
The developed stellar bar is found to possess  exponential density profile 
along the major axis of the stellar bar which is the so-called 
`exponential bar' (See \cite{el96} for a recent review.).
The initial disk is found to be heated dynamically in the vertical
direction of the disk  within  15 dynamical time owing to the dynamical
interaction between the background stars,  the small gaseous clumps, and
the developed stellar bar. Accordingly,  the present merger model describes 
the galaxy mergers between disk galaxies that $could$  be identified
with late-type barred galaxies $if$ galaxy merging had not been occurred.
Figure 2  is the time evolution of star formation
rate, $f_g$, for each merger model,
Model B1 and B5.
As is indicated in this figure, the star formation during galaxy merging
proceeds in such a way that the interstellar gas is consumed by star formation
more rapidly for the model with larger $C_{\rm SF}$.
This trend of star formation  holds for other model sequences Model B $\sim$  G.
The $C_{\rm SF}$ dependence of morphology and  dynamical structure in merger remnants
is briefly summarized in the Appendix section, which can help to understand more deeply
the present numerical results.
By using these models of galaxy mergers,
we  mainly present the $C_{\rm SF}$
dependence of 
mean stellar and gaseous metallicity, radial gradients of age and metallicity,
and the shapes of isophotes and isochromes,
in the following subsections.

\placefigure{fig-3}
\placefigure{fig-4}

\subsection{Dependence of the chemical properties on the rapidity of star formation}
Chemical, photometric, and spectroscopic
properties of galaxies depend basically on how
the metallicity and age of  stellar populations of galaxies
are distributed.
Thus, we begin to observe the distribution of  metallicity ($Z_{\ast}$) 
and the epoch of star formation ($T_{\ast}$)
in the stellar component of  
merger remnants and its dependence on  $C_{\rm SF}$. 
In this subsection \S 3.1, we present the following three results:
(1) Dependence of $Z_{\ast} - T_{\ast}$ distribution on $C_{\rm SF}$,
(2)  Dependence of mean $Z_{\ast}$ on $C_{\rm SF}$  and its physical reason,
and (3) The difference in the metallicity distribution  of merger remnants
between the Simple one-zone model and the present chemodynamical one.

\subsubsection{Distribution of stellar populations}

First we describe the results concerning the
dependence of $Z_{\ast} - T_{\ast}$ distribution on $C_{\rm SF}$. 
Figure 3  shows the distribution of stellar populations
on the $Z_{\ast} - T_{\ast}$ map 
for model with $C_{\rm SF}$ = 0.35  and for 3.5, respectively. 
The meaning of the $T_{\ast}$ is  that  the stellar component
with smaller $T_{\ast}$ is  born earlier thus $older$.
The vertical hight  in each bin of $Z_{\ast}$ and $T_{\ast}$ 
denotes the number of star particles with $Z_{\ast}$ and $T_{\ast}$. 
In this figure, we can observe that there is an appreciable scatter  in 
the metallicity for a fixed $T_{\ast}$.
This scatter of metallicity for a given $T_{\ast}$ 
 is one of the  characteristics of 
chemical evolution of galaxies in which 
chemical components produced by star formation are less efficiently  
and less widely mixed into the whole region of galaxies, 
as has been already pointed out by
Steinmetz \& M\"uller (1994) 
for  dissipatively collapsing galaxies.
$Z_{\ast}- T_{\ast}$ distribution in the model with $C_{\rm SF}$ = 3.5
is  shifted toward the larger $Z_{\ast}$  and smaller $T_{\ast}$ 
(older in age) than in the model with $C_{\rm SF}$ = 0.35,
meaning that a larger amount of metal-enriched stars are produced
more rapidly in the merger with rapid star formation
than in that with gradual one.
Spread in $T_{\ast}$ for a fixed $Z_{\ast}$ is more likely to be larger
in the model with $C_{\rm SF}$ = 0.35 than in  that 
$C_{\rm SF}$ = 3.5 
whereas $Z_{\ast} - T_{\ast}$  distribution is more strongly peaked
in the model with $C_{\rm SF}$ = 3.5 than in  that  
 $C_{\rm SF}$ = 0.35.
These results imply that in the model with $C_{\rm SF}$ = 0.35,
star formation proceeds less rapidly and more gradually,
resulting in a smaller amount of metal-enriched stellar populations
and a larger amount of younger stellar populations.
This  tendency of stellar populations
 plays an important role in determining the mean metallicity
and age, and thus the photometric and spectroscopic evolution of 
merger remnants, as is described later.

Second, we observe how the mean stellar metallicity depends on the
rapidity of star formation and explain the origin of the dependence.
As is shown in 
Figure 4, which  describes the time evolution of mean
stellar metallicity
($<Z_*>$) of each merger
remnant,  the chemical enrichment proceeds faster and 
more efficiently in the
model with larger $C_{\rm SF}$ than the  model with smaller $C_{\rm SF}$,
and thus the final value of $<Z_*>$  is 
larger in the model 
with larger $C_{\rm SF}$ than the  model with smaller $C_{\rm SF}$:
The time when $<Z_*>$ exceeds 0.015  is 
$T=0.7$ for  the model with $C_{SF}=3.5$
and  $T=5.0$ for the model with $C_{SF}=0.35$,
and the asymptotic value of $<Z_*>$ (the value at $T$ = 15.0 in our units)
is 0.0237 for the model with $C_{SF}=3.5$ and 0.0185
for the model with $C_{SF}=0.35$.
This result clearly demonstrates that in dissipative galaxy mergers
with star formation, the rapidity of star formation is an important
factor even for determining the global and average chemical properties 
of merger remnants.
The physical reason for this 
dependence can be explained   as follows.
For a galaxy merger with more rapid star formation, 
less  amount of interstellar gas 
is tidally stripped away
from the system 
during  galaxy merging 
principally because 
 more amount of initial gas has been already converted to stellar component
before the system suffers more severely from violently varying gravitational
potential of galaxy merging.
As a result of this,  more amount of the gas is consequently metal-enriched to
a larger extent 
and converted  into stellar component during merging.
Thus,
more amount of metals 
is shared by stellar component in
the  remnant of the  galaxy merger with 
more rapid star formation. 
Moreover this observed tendency of the $C_{SF}$
dependence of mean stellar metallicity is found to be applied to
the five standard models (See the seventh column in the 
Table 1 for Model B1 $\sim$ B5.).
These results provide a potential success of reproducing qualitatively
the mass-metallicity
relation of elliptical galaxies: If the  value of the $C_{SF}$ is larger for
more luminous galaxy merger, then the derived result indicates
that galaxy mergers between more luminous spirals are more likely
to become ellipticals with larger stellar metallicity.
This furthermore suggests that even the origin of fundamental 
photometric and spectroscopic  properties such as the color-magnitude relation
in elliptical galaxies are  closely associated with the details
of galactic dynamics. 

\placefigure{fig-5}

 In order to observe more clearly the above strong dynamical coupling
between dynamical evolution and chemical one, we perform a set of 
comparative simulations in which chemical evolution (including
star formation) is solved but the dynamical
one (gravitational and dissipative dynamics)
is not for an isolated galactic disk (Model A3 and A4).
Accordingly,  the density distribution
and kinematical properties in the disk are fixed during the simulations.
Since the time evolution of the gas mass fraction is the most
fundamental determinant for the mean stellar metallicity in galaxies,
by investigating the effects of galactic dynamics on the time evolution
of gas mass fraction and the final gas mass fraction,
we observe how the dynamical evolution can affect the chemical one
in the present merger model. 
Figure  5 gives the time evolution of gas mass fraction, $f_{\rm g}$,
for isolated disks $without$ $dynamical$
$evolution$ (Model A3 and  A4) and for mergers (Model B1 and  B5).
As is shown in the lower panel of
this Figure 5, the gas mass fraction at $T = 15.0$ is not so different
between models with different $C_{\rm SF}$ for isolated disks without dynamical
 evolution (Model A3 and A4) whereas significant difference in the final
gas mass fraction can be seen  
for mergers (Model B1 and B5).
This result clearly 
indicates  that dynamical evolution itself greatly strengthens the 
difference of final gas mass fraction 
between   
merger models with different $C_{\rm SF}$.
Therefore the chemical evolution in galaxy mergers 
is demonstrated to be
greatly affected by the dynamical evolution of galaxy merging. 
In the merger model (Model B1 and B5), as is described before, the total amount of 
interstellar gas tidally stripped away from the system 
is determined
by the rapidity of star formation ($C_{\rm SF}$), accordingly
the final gas mass fraction is strongly affected by the rapidity of star formation.
This sort of importance of the rapidity of star formation ($C_{\rm SF}$)
in controlling the final gas mass
fraction can be seen even in the isolated disk models (Model A1 and A2),
 and can be explained
as follows (See the upper panel of the Figure 5.).
In the model with smaller $C_{\rm SF}$, non-axisymmetric structures such as spiral
arms and bars developed in the star-forming disk can redistribute the mass
and angular momentum of the remaining gaseous component more efficiently
and for a longer times
principally because a larger amount of gas has not been yet  converted into
stellar component.
Consequently,  a larger amount of the gas is transported outwardly in the disk
and thus the mean gas density there is greatly dropped off.
As a result of this, only a smaller fraction of the outwardly transported
gas can be further converted into stellar component,
and thus the mean stellar metallicity is smaller for the model with smaller $C_{\rm SF}$.
This result clearly shows that the redistribution of mass
and angular momentum driven by  specific non-axisymmetric structures
plays an important role in determining the final gas mass fraction
in isolated disks.
Thus, although the dynamical mechanisms operating to yield the difference in
the final gas mass fraction between models
with different rapidity
of star formation is significantly different between isolated disks (Model A1 and A2)
and mergers (Model B1 and B5),
 however these results say  essentially the same thing:
Chemical evolution of galaxies can be greatly affected by 
mass and angular momentum redistribution
driven by global  dynamical  process such as the violent relaxation
of galaxy merging and mass-transfer triggered  by non-axisymmetric 
gravitational potential of galaxies.
This sort of strong coupling between dynamical evolution and chemical one 
is first demonstrated by the present work and Bekki \& Shioya (1997c).

\placefigure{fig-6}

 Third, we observe the difference in chemical evolution of merging galaxies
between the Simple one-zone model and the present chemodynamical one.
Figure 6 describes the metallicity distribution of stellar component for
the model with $C_{\rm SF}$ = 0.35  and for 3.5,     
on each of which the metallicity distribution predicted from the Simple
model is superimposed.
In estimating  the distribution of stellar metallicity expected from 
the Simple model for each model,
we use the time evolution of gas mass fraction obtained for each
merger model (Model B1 and B5),
and then calculate the metallicity distribution
according to the standard formulation
on the relation between gas mass fraction and mean stellar metallicity
for the Simple model. 
Since the Simple model assumes both instantaneous recycling and instantaneous
mixing, the Simple model corresponds to the present chemodynamical model with
$R_{\rm che}$ equal to infinity.
We also give the values of mean stellar and gaseous metallicity expected from
the Simple model in the ninth and tenth column of the Table 1, by adopting
the following formulation: 
\begin{eqnarray}
<Z_{\ast}> & = & 
y_{\rm met} \left( 1 + \frac{f_g \ln f_g}{1 - f_g} \right) \; \; \; ,\\
Z_{\rm gas} & = & 
- y_{\rm met} \ln f_g \; \; \; .
\end{eqnarray}

As is shown in Figure 6, the metallicity distribution of the model with
larger $C_{\rm SF}$ is shifted to
righter side than the model with smaller $C_{\rm SF}$,
which is consistent with the already obtained trend that the mean stellar
metallicity of merger 
remnants with larger $C_{\rm SF}$ is larger than that of the remnants 
with smaller $C_{\rm SF}$.
Furthermore, it is found that irrespectively of the values
of $C_{\rm SF}$, the metallicity distribution is shifted to the righter 
direction in the chemodynamical model than in the Simple
one and thus that the spread of the distribution
is larger in the chemodynamical model.
This larger spread results from the fact that in the present 
chemodynamical model, the chemical enrichment proceeds preferentially
in the gaseous region with  higher density
where a larger amount of chemically enriched gas is converted 
into new and more metal-enriched stars.
The magnitude of the difference of the distribution
is found to be more likely to be larger for models with smaller $C_{\rm SF}$. 
This result clearly explains the reason why the 
difference in the mean stellar metallicity   between the  models with
different  values of $C_{\rm SF}$ for the present chemodynamical model
is smaller than expected from the Simple model (See the sixth,
seventh, ninth, and tenth column in the Table 1.).
Gaseous metallicity, on the other hand, is found to be
smaller in the chemodynamical
model compared with that expected from the Simple model and not to
depend on the values of  $C_{\rm SF}$. 
These differences in the metallicity distribution between the Simple model
and chemodynamical one results in the difference between these model
in the photometric and spectroscopic properties of merger remnants,
as is described in detail later.

Thus we demonstrate that dynamical evolution of galaxy merging greatly affects
the chemical evolution in galaxy mergers, and consequently the
chemical evolution in mergers is significantly different from that
expected from the Simple one-zone model.  
What is  particularly important in the present study
is that the tidal stripping of interstellar
gas during galaxy merging play a vital role in determining even the global
chemical  properties of merger remnants.
This tidal stripping of interstellar gas is probably more important
in galaxy mergers with a larger amount of interstellar gas, for example,
in the mergers at higher redshift or in the mergers between less luminous
spirals.

\placefigure{fig-7}
\placefigure{fig-8}

\subsubsection{Radial  gradient of stellar populations}
 Here, we first observe how successfully the observed profile of
radial metallicity  gradient of
elliptical galaxies can be reproduced in merger remnants and
how the rapidity of star formation controls the absolute magnitude
of the gradient. 
Observational studies have revealed that the radial metallicity gradient of 
 elliptical galaxies can be fitted reasonably well by the power-low profile
and that the magnitude of the gradients, 
$\Delta \log Z_{\ast}/\Delta \log R$, are -0.2 on average 
(Peletier et al. 1990; Davies et al. 1991).
Mihos \& Hernquist (1994) found that the collisionless galaxy mergers
between spirals with exponential profile of  metallicity gradient 
become ellipticals 
with the metallicity gradient profile appreciably  deviating from
power-low profile,
implying that the later mergers with less amount of gaseous dissipation
can not reproduce the observed gradient with  power-low profiles.
Figure 7  gives the radial distribution of stellar  and gaseous 
metallicity for the
present merger model with star formation and  gaseous dissipation.
As is shown in this figure, negative metallicity gradient fitted by
power-low profile is reproduced reasonably well for model with
$C_{\rm SF}$ = 0.35 and 3.5, which indicates that the origin
of the metallicity
gradients of elliptical galaxies 
are closely associated with  dissipative
galaxy merging with star formation. 
A merger remnant in the  model with $C_{\rm SF} = 0.35$ has
larger stellar metallicity in the central part and  smaller metallicity
in the outer part than the remnant in the model with $C_{\rm SF} = 3.5$,
 meaning  that the metallicity gradient is larger for the model
with less rapid star formation.
The qualitative explanation for this trend of metallicity gradients
can be given as follows.
In the model with smaller $C_{\rm SF}$, a larger amount of 
metal enriched gas can be transferred to the inner region of
the remnant owing to less rapid star formation and more amount of gaseous 
dissipation.
In the outer part of the merger with  smaller $C_{\rm SF}$,
 star formation
is less likely to proceed efficiently and thus less likely
to form metal-enriched stars
because of the lower gaseous density in the outer part.
As a result of this, a relatively larger  metallicity gradient
is established for the model with smaller $C_{\rm SF}$.
For the model with larger $C_{\rm SF}$,
on the other hand, a larger amount of gas has been converted into
metal-enriched stars before the onset of violent relaxation,
thus the stellar component in the  galaxy merger suffers more severely from
chemical mixing driven by violent relaxation owing to less amount of
gaseous dissipation.
As a result of this, the metallicity gradient is more strongly washed out
and consequently the absolute magnitude of it becomes smaller for model with larger 
$C_{\rm SF}$. 
The obtained  value of metallicity gradient 
 is found to be smaller than that of 
the typical value of observation, -0.2, for most   of models with $R_{\rm che}$ = 0.4.
This result indicates that in the present chemodynamical model of
galaxy mergers, chemical mixing, controlled basically
by the chemical mixing length and the degree of violent
relaxation of galaxy merging,  is appreciably more efficient
than required for reproducing the observed trend.

Next we observe the dependence of the radial gradient of stellar age on
the rapidity of star formation.
It has been generally considered that
it is difficult to derive the age gradient observationally, 
since the color gradient is affected both by  metallicity gradient 
and by age gradient (i.e., the  age-metallicity degeneracy problem).
Recently, Faber et al. (1995) and Bressan et al. (1996) have
succeeded in breaking  the degeneracy 
by using the combination of line strength indices, H$\beta$ and [MgFe].
Faber et al. (1995) furthermore 
found that outer parts of elliptical  galaxies are more likely to be
both older 
and more metal-poor than nuclei. 
This observational trend is found to be reproduced more successfully
in the model with smaller $C_{\rm SF}$ for the present merger model,
suggesting that such a trend  of  age and metallicity
gradient of elliptical galaxies 
is just one of the evidences that an elliptical galaxy is  formed by
galaxy merging.
Figure 8 describes the radial gradient of mean epoch of
star formation, $<T_{\ast}>$, which corresponds to the age gradient
in merger remnants.
As is shown in Figure 8, 
the age gradient is more discernibly seen in the model with 
$C_{\rm SF}$ = 0.35 than the model with $C_{\rm SF}$ = 3.5,
 primarily because in the model with $C_{\rm SF}$ = 0.35,
the star formation proceeds more gradual and more later
in the inner region of the merger remnant.
Since it is found that the stellar metallicity is larger for the central
region than for the outer region of a merger remnant, 
this result   means that in the model with $C_{\rm SF}$ = 0.35,
the inner part of the merger remnant show  both younger age and a larger
degree of chemical enrichment whereas 
in the model with $C_{\rm SF}$ = 3.5, the inner part of the merger remnant show
both slightly older age and a slightly larger  degree of chemical enrichment. 
Interestingly, the result for the model with $C_{\rm SF}$ = 3.5 
can be seen also in the result of
Bressan et al. (1996), which  describes  that 
there are a number of  elliptical galaxies whose nucleus are both older and 
more metal-enriched. 
Thus, the rapidity of star formation is demonstrated to determine
both the profiles and the magnitude of the metallicity (and age) gradients
in galaxy mergers:
Both the metallicity gradient and the age one in the model with rapid star formation 
are  
shallower than that with less rapid star formation.

\placefigure{fig-9}

\subsubsection{The shape of isochromes}

The shape of isochromes and the degree of the difference
between the isochromes  and isophotes shapes are
considered to be  observational clues
about the origin of the formation and evolution
of elliptical galaxies.
As is revealed by the  
observation of multi-band photometry 
(e.g. Boroson et al. 1983),
the shapes of isochromes 
are appreciably similar to those of isophotes in  elliptical 
galaxies. 
 In the dissipative collapse
model of Larson (1975), the shapes of isochromes are
 significantly  flatter than those of isophotes,
which is not in agreement with the above observation. 
In the Carlberg (1984)'s dissipative
collapse model of elliptical galaxy formation,
 the  shapes of isochromes are
slightly flatter than  
the isophotal shapes and thus appears to have succeeded in overcoming
the disadvantage of the Larson's collapse model regarding the
difference between isochromes and isophotes shapes.
Here, we observe whether or not
the shape of isochromes can  match reasonably well with that of
isophotes in the present dissipative merger model. 
Although the color of a galaxy is affected by both metallicity and age 
of stellar populations, 
we assume  that the iso-metallicity contour is exactly the same
as  the isochromes one 
and that  the iso-density contour is also the same as  the isophote
one. 
In Figure 9, we show  arbitrary isochromes (solid line) and 
isophotes (dotted line) for the model with $C_{\rm SF}$ = 3.5. 
We found that the shapes of isochromes are  nearly the same as 
those of isophotes 
and furthermore that this  trend is not  dependent so strongly on the 
$C_{\rm SF}$. 
These results suggest that violent relaxation combined with an appreciable
amount of gaseous dissipation,
which is a specific dynamical process for dissipative galaxy 
mergers,  plays a vital role in
producing the shape of isochromes which can match well with the 
isophotal shape.

\placefigure{fig-10}
\placefigure{fig-11}
\placefigure{fig-12}
\placefigure{fig-13}

\subsection{Fundamental  roles of  other parameters}
In addition to the rapidity of star formation,
chemical mixing length, the orbit configuration of galaxy merging,
the multiplicity of a merger, and the epoch of galaxy merging
are considered to be  fundamental factors for chemical, photometric,
and spectroscopic properties. 
We accordingly 
observe how and to what extent the above four parameters strengthen or lessen
the $C_{\rm SF}$ dependence of chemical properties derived in \S 3.1. 
Dependence of chemical properties on the four parameters
is  summarized in 
Figure 10 for mean stellar metallicity,  Figure 11 for mean epoch of star formation,
Figure 12 for stellar metallicity gradient, Figure 13 for radial gradient of
mean epoch of star formation.
What should be emphasized in these figures is that
although variety in the values of the above  four parameters
indeed introduces  appreciable  
scatter in the $C_{\rm SF}$ dependence of chemical properties derived in
\S 3.1, however, the basic trends in the $C_{\rm SF}$ dependence 
derived in \S 3.1 is
not so drastically changed.  
For example, the mean  stellar metallicity 
of  a merger remnant is larger in  the model 
with larger $C_{\rm SF}$ for a give chemical mixing length,
orbit configuration, multiplicity,
and merging epoch.
Accordingly, in the following parts (\S
3.2.1 $\sim$ 3.2.4),  we mainly present outstanding
results which we could  not obtain 
only by varying the value of $C_{\rm SF}$
in the previous subsection \S 3.1,
and furthermore  we describe the fundamental roles
specific for the above four parameters in determining the chemical
properties of merger remnants.

\subsubsection{Mixing length of chemical component}
 It is considered to be highly uncertain how locally  
and how well    metals ejected from stars with different masses
and ages can be actually mixed into interstellar medium in galaxies 
(e.g., Roy \& Kunth 1995).
Because of this uncertainty, it is the  best way in the preliminary
stage of the present study to vary the values of the chemical mixing
length, represented by $R_{\rm che}$, in each model 
 and thereby to examine its importance in determining chemical properties
of galaxies. 
We found that irrespective of the values of $C_{\rm SF}$,
as the values of $R_{\rm che}$
becomes smaller,  
the mean stellar metallicity becomes larger (See the seventh column
in the Table 1 for Model
B6, 7, 8, and 9). 
This result can be explained as follows.
In the model with smaller $R_{\rm che}$,
 metals  produced by star formation 
in higher density region can  be mixed into
the surrounding  interstellar gas more locally,  
and accordingly  the  metals 
are more likely to be trapped only  by the gas in the higher density region. 
Consequently, chemical enrichment proceeds more  preferentially
in  the  higher density regions, where a larger number  of stars are 
formed 
and thus the mean stellar metallicity in the merger is basically determined.
As a result of this,
a larger number of stars 
are more likely to be born from the gas with a larger metallicity, 
and thus the mean stellar metallicity of merger remnants is larger
for galaxy mergers with smaller $R_{\rm che}$.
This result indicates that if the $R_{\rm che}$ is larger for 
the model with smaller $C_{\rm SF}$, the already obtained 
$C_{\rm SF}$ dependence of mean stellar metallicity 
is strengthened (See Figure 10.).
We also found that for most models,
as the values of $R_{\rm che}$ is smaller,
the negative metallicity gradient becomes steeper (See Figure 12).
This result reflects the fact that the chemical enrichment 
can proceed more preferentially  in the higher density (i.e., inner)  region
such as the nuclei of mergers in the model with smaller $R_{\rm che}$.

 What is the most interesting result derived by varying the $R_{\rm che}$
is that for the model with smaller $R_{\rm che}$ and $C_{\rm SF}$,
the mean gaseous metallicity is smaller than
the stellar one in the merger remnant (See the sixth and seventh
column in the Table1 for Model B6 and B8).
This result can not be
obtained until both dynamical and chemical evolution are solved in
a admittedly self-consistent manner, 
since the Simple one-zone model predicts that the gaseous metallicity is
never smaller than the stellar one.
Furthermore, this result provide a valuable clue about the
origin of metal-poor hot gaseous halo of elliptical galaxies:
The $X$-ray satellite $ASCA$ 
($Advanced$ $Satellite$ $for$ $Cosmology$ $and$ $Astrophysics$), the metallicity
of the hot $X$-ray gaseous halo is appreciably smaller than
that of the stellar component of the host elliptical galaxy (e.g.,
Matsumoto et al. 1997).
The details of the formation of metal-poor gaseous halo will be discussed in our
future papers.

\subsubsection{Orbit configuration}
 Orbit parameters of galaxy merging such as the initial intrinsic
spin of progenitor disks and pericenter distance of mergers can affect
the degree of violent relaxation and the redistribution
of mass and angular momentum during merging, thus determine
the transfer process of
chemical component in mergers.
We accordingly investigate how the initial inclinations of two progenitor
disks (Model C, D, and  E) and pericenter distance (Model F) can affect the final 
chemical properties of merger remnants.
We found that for the model with larger $C_{\rm SF}$ in each model sequence, 
Model C, D, E and F, 
the mean stellar metallicity of merger remnants is larger
whereas the mean epoch of star formation ($<T_{\ast}>$) is smaller 
(See Figures 10 and 11.), 
which is consistent with the results derived in \S 3.2 
and thus reinforces the importance of the rapidity of star formation
in determining the mean stellar metallicity and mean stellar age  in merger remnants.
As is shown in Figure 12,
both the magnitude of negative metallicity gradient 
and that of age gradient in merger models
are distributed with  appreciable spread
even for a fixed value
of $C_{\rm SF}$ (=0.35 and 3.5), suggesting that the spread in 
the magnitude of metallicity 
and age gradient observed in  elliptical galaxies with a given
luminosity is due to
the diversity in orbit configuration of galaxy merging. 
The origin of the relatively larger negative value of metallicity gradient
observed for model with nearly prograde-prograde merger (Model D2) is
probably that the developed prolate stellar bar in the SB0-like merger
remnant more efficiently transfers the chemical component into 
the central region, 
and consequently enhances the difference in stellar metallicity
between  outer and inner region in the galaxy. 
This result should be compered with the result of Friedli \& Benz (1995)
 in which
the rotating stellar bar changes the mass and angular momentum
distribution of stars and gas in a galactic disk, and consequently
smoothes out the already existing metallicity gradient of the 
barred galaxy. 
The difference in the dynamical roles of stellar bars between
the present study and that of  Friedli \& Benz (1995)
 probably results from the differences in the 
total amount of gas mass and the strength of dynamical
perturbation assumed in the two studies.
More detailed studies are required for understanding more clearly the 
dynamical roles
of triaxiality (i.e., barred structure) in determining the chemical gradient of
galaxies and their dependences on galaxy types. 

\subsubsection{Multiple galaxy merging}
 Multiple mergers between late-type spirals are suggested to 
play important roles, for example,  in  forming  field elliptical galaxies 
(Barnes 1989)
and in mitigating  the difficulty of pair mergers to
produce the observed smaller degree of kinematical misalignment (Weil \&
Hernquist 1996).
We here present  the difference in a number of
chemical properties of merger remnants between  pair mergers and
multiple ones (Model G1 and G2).
The differences obtained in the present study are the following four:
First, irrespectively of
$C_{\rm SF}$, 
a larger amount of interstellar gas is stripped away and consequently
can not participate further star formation in  multiple mergers than
in pair mergers. This is
firstly because the tidal stripping of interstellar gas is more efficient in 
multiple mergers owing to  the more violent dynamical interaction between
galaxy mergers and secondly because a smaller amount of mass
can be transferred to the inner region and thus  can  not form
the higher density gaseous regions in  multiple mergers.
Second, as a result of this, a larger amount of metal-poor gas 
 finally surrounds  the merger remnant in  multiple mergers,
which might be observed as the metal-poor gaseous $X$-ray halo as is actually
observed in elliptical galaxies (e.g., Matsumoto et al. 1997).
Third, the  $C_{\rm SF}$ dependence of chemical properties in pair merger remnants,
in particular, the $C_{\rm SF}$ dependence of  metallicity
gradient,  are less discernable in multiple mergers.
This is probably 
because chemical mixing driven by violent gravitational relaxation
of galaxy merging is more effective in multiple mergers.
Fourth, in the multiple merger model with  $C_{\rm SF}$ = 3.5,
the stellar populations in the central part of the merger remnant
is both older and less metal-enriched than the outer part,
which is a specific characteristic of chemical properties 
for multiple mergers with larger  $C_{\rm SF}$ in the present study.
This result is reflected in  the fact that in the multiple
merger  model with $C_{\rm SF}$ = 3.5, the magnitude 
of metallicity gradient for the
region,  $0.5 R_{\rm eff} \leq R \leq 5.0 R_{\rm eff}$ ($R_{\rm eff}$ is effective
radius.), is -0.07 whereas it is 0.19 for the region,
 $0.1 R_{\rm eff} \leq R \leq 1.0 R_{\rm eff}$.  
Actually, Bressan et al. (1996)  show that there are a number of galaxies 
possessing the same trend as is described above.  
Since only a relatively smaller range of parameter space has been investigated 
in the present study, we could not here give a stronger statement that
the elliptical galaxies with older and less metal-enriched stellar
populations in the central part of the galaxies are indicative of 
multiple mergers.
Accordingly, it would be our further study to confirm the above result
by performing numerical simulations
with a larger range of parameter space for multiple mergers
and furthermore to investigate what is the fundamental 
chemical properties specific for multiple mergers.
Thus, these four results demonstrate that the multiplicity in
galaxy mergers can also affect a number of chemical properties in 
merger remnants (thus in elliptical galaxies).

\placefigure{fig-14}
\placefigure{fig-15}

\subsubsection{Epoch of galaxy merging}
 We here investigate how the difference in the epoch of galaxy
merging can affect the fundamental chemical properties of  merger remnants
by varying the initial separation between two progenitor disk ($r_{\rm in}$)
and
thereby delaying the epoch of galaxy merging (Model B10 $\sim$ 15).
We found that a galaxy merger
 with larger $r_{\rm in}$ (later merger) becomes an elliptical galaxy with both 
more metal-enriched and younger stellar populations
and furthermore that this tendency does not 
depend on the values of $C_{\rm SF}$ (See the Figure 14.). 
These results can be explained as follows:
In the later galaxy merger (the model with larger $r_{\rm in}$),
a larger amount of interstellar gas can continue to be  converted into stellar
component  and chemically
enriched in the progenitor  disk for a longer time,
since the merging epoch is delayed.
Consequently, 
new stars (younger stellar populations)
are more preferentially born from  
more chemically enriched interstellar gas 
when the later starburst is triggered by the later galaxy merging.
As a result of this, stellar populations in the later
galaxy merger are both younger and more metal-enriched on average
than the earlier merger.
Faber et al. (1995) and 
Worthey et al. (1996) suggest  that if the younger galaxies are actually  more 
metal-rich than older ones, the color-magnitude relation of elliptical galaxies
can be equally reproduced without invoking the conventionally believed
mass-metallicity relation of elliptical galaxies.
They furthermore suggest that 
even if there exists  outstanding spread in the luminosity-weighted age
and metallicity among elliptical galaxies,
the tightness of the color-magnitude relation can be maintained
with a specific assumption that  the age and metallicity, $Z$, 
satisfies the relation,  $\Delta \log {\rm age} / \Delta \log Z$ = -1.5
(Worthey's law).
It is remarkable  that the above numerical result that a merger remnant 
with younger stellar populations is more metal-enriched is
at least qualitatively consistent with the Worthey's law.
Although more extensive observational studies should be accumulated which can
confirm whether or not the  Worthey's law is universal among elliptical
galaxies with different luminosity and with different environment,
we can say however that the nature of  stellar populations of elliptical
galaxies is closely associated with the epoch of galaxy merging
and  thus with the strength of the secondary burst of star formation
during galaxy merging.
Specifically, it is crucial for chemical and photometric properties
of elliptical galaxies formed by galaxy merging how much
amount of interstellar gas  the progenitor disks
still have and to what degree the gas has been already metal-enriched
just before galaxy merging.
Although our numerical study appears to have succeeded in grasping 
some essential ingredients concerning the dependence of the chemical
properties of merger remnants on the epoch of galaxy merging,
a larger range of parameter space should be covered by our future
studies in order to confirm the derived dependence.

Finally in this subsection \S 3.2, we give the dependence of age gradient on metallicity one
for all merger models in Figure 15.
As is shown in this Figure 15, the merger model with larger magnitude of metallicity gradient
show the larger magnitude of age gradient, which is actually observed in Bressan et al. (1996).

\placefigure{fig-16}
\placefigure{fig-17}

\subsection{Characteristics of stellar
populations dependent on galactic luminosity}

We have so far focused mainly on
the parameter dependence of chemical properties of merger 
remnants  and did not intend to describe the photometric and spectroscopic
properties of the remnants by adopting a specific mass and size
for each merger model
in the previous subsections (\S 3.1 and 3.2), 
We here describe photometric and spectroscopic properties of
merger remnants by assuming that the model with $C_{\rm SF}$ = 1.0
corresponds to the galaxy merger between late-type spirals with
mass and size equal to the fiducial values of the present study
($M_{\rm d}$ = 6.0 $\times$ $10^{10}$ $ \rm M_{\odot}$ and
  $R_{\rm d}$ = 17.5 kpc, respectively) and that $C_{\rm SF}$
$\propto$ $L^{0.55}$
(This assumed relation is plausible, see the Appendix A.).
The assumed relation, $C_{\rm SF}$ $\propto$ $L^{0.55}$, means
that more luminous elliptical galaxies are formed by galaxy merging
with more rapid star formation.
For convenient, we consider that for each model, 13 Gyr have 
passed since the two progenitor galaxies began to merge 
with a given $r_{in}$.
This means that we do not intend to include any $initial$ age difference between
galaxy mergers in this consideration,
thus means that the derived photometric and spectroscopic
properties basically
reflect the mean stellar metallicity of the remnants. 
It should  be emphasized here that whether or not the above
assumptions on the scaled mass and size of galaxy mergers are 
$actually$ reasonable
and realistic is probably highly uncertain in the present
study because of the lack of extensive observational studies on the luminosity
dependence of star formation histories of disk galaxies and mergers.
Accordingly, the derived luminosity dependence of photometric 
and spectroscopic properties
of merger remnants only reflect the adopted assumptions,
and thus we describe them  only in a
schematic manner.
Although this investigation could not allow us to compare the photometric
properties of merger remnants with the observational ones of real elliptical
galaxies in a more quantitive way, however, it enable us to point out 
in a quantitative way
the possible advantages and disadvantages of the present merger model in
reproducing the observed photometric properties of real elliptical
galaxies. In the following parts (\S 3.3.1 and 3.3.2), 
luminosity dependence of mean stellar metallicity,
metallicity  gradient, and integrated color
of merger remnants are mainly presented.

\subsubsection{Chemical properties}
Based upon the results presented in the Figures 10, 11, 12, and 13,
we can obtain the following dependence:
More luminous elliptical galaxies formed by galaxy merging
show (1) larger mean stellar metallicity, (2) smaller central stellar metallicity,
and (3) smaller radial gradient of stellar metallicity.
The first result appears to agree qualitatively with the mass-metallicity relation
implied by the color-magnitude relation of elliptical galaxies
(Faber 1973; Visvanathan \& Sandage 1977), whereas the second result
appears to disagree with the expected luminosity dependence of the 
central metallicity of elliptical galaxies
indicated by the $\rm Mg_{2}$ - $\sigma$ relation (Burstein et al. 1988).
It still seems less feasible  to  determine
whether the third  result matches with the observed luminosity 
dependence of stellar metallicity gradient in elliptical galaxies, as is
described below.
Although a number of observational studies have been accumulated
which describe the dependence of the metallicity gradient on
galactic parameters, however, the clear trend in the dependence
seems less likely, as is described below. 
Carollo et al. (1992) reported that Mg$_2$ index 
gradient shows a bimodal trend with mass: 
For less massive galaxies ($M < 10^{11} M_{\odot}$), 
the Mg$_2$ gradient increase with increasing mass whereas 
for massive galaxies ($M > 10^{11} M_{\odot}$), 
there is no obvious pattern in mass dependence of the gradient. 
Gonzalez \& Gorgas (1996) found the gradient of Mg$_2$ index correlates 
with  the central Mg$_2$ index:
Galaxies with  steeper gradient of Mg$_2$ index have 
larger central Mg$_2$ index. 
Since central Mg$_2$ index becomes larger as the mass of galaxy increases, 
this correlation suggests  
that massive galaxies are more likely to have steeper metallicity gradient.  
Furthermore, Peletier et al. (1990) reported that there is no correlation between 
color gradient, which is a combination of age and metallicity gradients,
and the luminosity of galaxies. 
Thus, it is safe for us to say  that
we could not here give any implications on the origin for  the dependence
of metallicity gradient on the galactic parameters.
We must wait further extensive studies which reveal  clear trends in
the dependence
of metallicity gradient on the galactic parameters, in order to compare 
the derived  $C_{\rm SF}$ dependence of metallicity gradient in the
present study with the observational results.

\subsubsection{Photometric and spectroscopic properties}
 Figure 16 is the radial distribution of $U-R$ and $B-R$ 
color in the merger remnant for the fiducial model with  $C_{\rm SF}$ = 1.0. 
The observed color gradient in this model is principally due to
the metallicity gradient of the merger remnant, since the age difference
between outer part and inner one in the merger remnant is less than 1 Gyr.
We found that
this relatively shallower color gradient is  more discernibly
seen in more luminous merger remnants (mergers with rapid star formation).
Figure 17  describes the color-magnitude (CM) relation derived for 25 models.
In order to observe the difference in the photometric properties (mean color)
between the Simple one-zone model and the present chemodynamical one,
we calculate the photometric properties expected from the Simple model for the five 
standard  models
(Model B1 $\sim$  5) by using the star formation history obtained  for 
the five models. 
The derived results of the Simple one-zone model are plotted in the same 
figure for five models.
As is shown in Figure 17,  more luminous merger remnant (corresponding to
the galaxy merger with larger $C_{\rm SF}$) show the redder color 
in the $V-K$, however, the derived luminosity dependence
of the integrated color does not agree reasonably well with the observed
color-magnitude relation of elliptical galaxies.
What this result actually means depends on whether or not 
we adopt the conventional point of view regarding the
origin of the observed CM relation of elliptical galaxies.
In the conventional points of view (Faber 1973; Visvanathan \& Sandage 1977), 
the CM relation  reflects the mass-metallicity relation
of galaxies (`metallicity effect' ) whereas in the newly proposed view 
(Faber et al. 1995; Worthey et al. 1996),
the CM relation reflects  the fact that
less luminous elliptical galaxies are progressively 
younger than  more luminous ones (`age effect').
If we adopt the former conventional point of view, we should 
conclude as follows.
The above result  means that 
the observed color-magnitude relation (mass-metallicity relation)
could not be reproduced so successfully by 
the present merger model with  
$C_{\rm SF}$ $\propto$ $L^{0.55}$.
As has been already pointed out by Bekki \& Shioya (1997c),
even the present merger model, in which any age difference between
galaxies and any thermal and dynamical feedback effects of
star formation are not included, can reproduce at least qualitatively
the observed CM relation if we assume that
$C_{\rm SF}$ $\propto$ $L^{1.69}$.
However, this assumption  $C_{\rm SF}$ $\propto$ $L^{1.69}$
seems less plausible, since, for deriving the
$C_{\rm SF}$ $\propto$ $L^{1.69}$, 
we must adopt somewhat  particular
assumptions, for example, that the initial  density of
the progenitor disk ($\Sigma$) in a merger 
depends strongly on the galactic luminosity ($L$)
in such a way  that $\Sigma$ $\propto$ $L^{4.56}$. 
Therefore, instead of adopting such apparently less realistic
assumptions,
we should reconsider other fundamental parameters that are important
determinant for chemical evolution of galaxy mergers
and strongly depend on 
the galactic mass, and then again investigate the luminosity
dependent chemical and photometric properties of merger remnants.
On the other hand, if we adopt the latter fresh point of view 
(Faber et al. 1995; Worthey et al. 1996), 
we should conclude as follows.
The derived result on the CM relation  does
not necessarily disagree with the 
observed CM  relation of elliptical galaxies.
In this case, we must wait further extensive observational studies
which reveal to what extent each  `age effect' 
and `metallicity effect' actually contributes to  the generation 
of the CM  relation
and  clarify the relative importance of the two effects in
the reproduction of  the  CM relation.
It would be our future studies to elucidate 
the origin of the relative importance
in the context of dissipative galaxy merging,
when the relative contribution of the two  `age effect' and 
`metallicity effect' are observationally clarified. 
Considering these situations, it is  safe for us to say that
we can  not determine how successfully the present merger model
has reproduced the CM  relation until
more extensive observational studies clarify the
relative importance of the two
`metallicity effect' and `age effect'.

Lastly, we give a comment on the difference in
the photometric and spectroscopic properties between the Simple one-zone
model and the present chemodynamical one.
As is shown in Figure 17, the $V-K$ color is bluer in
the Simple model than in the chemodynamical one,
which results from the fact that
the stellar metallicity
in the Simple model is smaller compared with that of
the chemodynamical one.
This result implies that the chemical,  photometric, and spectroscopic
properties derived by adopting the Simple one-zone 
model is not better approximation as is tacitly understood, 
especially when the more quantitative comparison
of the theoretical results with observational ones are required.

\section{Discussion}

\subsection{Outstanding difference between 
the Simple one-zone model and the present chemodynamical one} 
 We here present the two outstanding differences in chemical
evolution between the Simple one-zone model and the present chemodynamical one,
and then give the related implications on the chemical
evolution of  elliptical galaxies formed by galaxy mergers.
First, we stress that even if we discuss the global and mean properties
of chemical and photometric evolution of elliptical galaxies formed by galaxy
mergers,
both chemical and dynamical evolution should be solved in an 
admittedly self-consistent way.
As is described in  detail in the previous sections,
chemical enrichment process
during  dissipative
galaxy merging with star formation, which is controlled  basically
by the violent relaxation of galaxy merging and gaseous dissipation,
is found to proceed in a considerably inhomogeneous way.
This nature of chemical enrichment process
provides  a remarkably difference between the Simple one-zone
model and the present chemodynamical one in that  the final gas mass fraction
is not only the important
determinant for the final stellar metallicity in the present chemodynamical model.
As a natural result of this difference,
photometric and spectroscopic
properties such as the integrated color in $V-K$ is observed
to be appreciably different between the Simple one-zone model and the present
chemodynamical one.
Although the difference in the photometric and spectroscopic properties
of merger remnants  between the two models 
is not so large, however, we should recognize how the 
details of dynamical evolution of forming elliptical galaxies modify the results
derived by the Simple
one-zone model, especially when comparing the theoretical results
more precisely with the observational ones.
Furthermore, in the present chemodynamical model,
chemical enrichment is found to proceed faster in the higher density
region (i.e., inner region) where star formation occurs more efficiently
than in the lower  density one (i.e., outer region), which strengthens the difference
in the degree of chemical enrichment between the higher  density 
(inner) region
and the lower (outer) one  
(`chemical segregation').
This sort of chemical segregation could 
give a natural explanation for  
the origin of the difference in the  photometric and spectroscopic properties between
the outer   and the central part in an elliptical galaxy.

Second, we should emphasize the fundamental  roles of 
chemical mixing length in determining
a number of chemical  properties
in merger remnants, such as the mean stellar metallicity and radial  metallicity 
gradient. 
Unlike the Simple one-zone model,
chemical enrichment is assumed to proceed $locally$ in the present merger model.
As is described in detail in the previous sections,  both radial metallicity
gradient and  mean stellar metallicity  in merger remnants depend on
how locally   metals ejected from stellar component can be
mixed into interstellar gas: Both mean stellar metallicity 
and the absolute magnitude of
radial gradient of stellar metallicity are larger for the merger remnant
with smaller mixing length of chemical component. 
Moreover, the chemical mixing length is found to  determine how much amount of
heavy elements ejected from stellar component can be shared by stars or interstellar
gas, accordingly to affect the formation of gaseous halo which is less 
metal-enriched than stellar component of merger remnants.
These results suggest that the rapidity of star formation in mergers 
is not only
the dominant factor which determines the chemical 
properties of merger remnants, thus  that we should investigate the 
physical relation between the rapidity of star formation
and chemical mixing length in galaxies and its dependence on galactic
luminosity for more clear understanding of the chemical, photometric,
and spectroscopic evolution of galaxy mergers.
There still remains a wide room for investigating and understanding
the chemical evolution 
of galaxy mergers.

\subsection{On the origin of color-magnitude relation}
 The color-magnitude (CM) relation of
 elliptical galaxies is considered to be one of the most fundamental
relations  containing 
valuable information about formation history of elliptical galaxies.
The CM relation describes that 
the integrated color of elliptical galaxies becomes 
redder as the absolute magnitude of the luminosity decrease (e.g., Faber 1973).
Although the age difference between elliptical galaxies has been recently demonstrated
to play an important role in the reproduction of the CM relation (e.g., Worthey et al. 
1996), it has been conventionally considered that 
the mean stellar metallicity is 
larger for more luminous elliptical galaxies (e.g., Tinsley 1978). 
Larson (1974) originally proposed that this luminosity-dependent
mean stellar metallicity observed in elliptical
galaxies is essentially ascribed to when
 star formation is entirely truncated owing to 
the so-called `galactic wind' 
driven by accumulated thermal energy
produced mainly by supernovae in the early stage of dissipative collapse. 
Although there is a large number of studies to adopt this `galactic wind'
model and to investigate the origin of the CM relation by using
evolutionary method of population synthesis (e.g., Arimoto \& Yoshii 
1987, more recently, Bressan et al. 1994),
however,  only a few studies have addressed  the fundamental question about
wether or not the merger model of elliptical galaxy formation
can also reproduce the mass-metallicity relation.
Thus, we begin to  discuss this important issue in terms of elliptical
galaxy formation by galaxy merging, provided that the color-magnitude relation
reflects the mass-metallicity relation in elliptical galaxies.

 As is described in the previous section, what is remarkable in the present 
merger model is that even if we do not include the thermal and 
dynamical feedback effects of type II
supernovae on the gas dynamics in merging galaxies, 
final gas mass fraction and mean stellar metallicity
in merger remnants depend on the star formation history and thus on galactic
luminosity owing to the tidal truncation of star formation.
This mechanism that the mass-metallicity relation can be closely associated
with the details of dynamics of galaxy merging is quite different from
that invoked by the classical galactic wind model.
This result  furthermore implies that there can be  a number of ways
to reproduce the mass-metallicity relation in elliptical galaxies
and thus  the  CM relation 
probably has more profound meanings than we can
deduce by invoking  only  a specific model of  elliptical galaxy formation. 
Here we should note that although our merger model have a potential success
in reproducing at least qualitatively the mass-metallicity relation,
more totally  and successful  comparison of our merger model with 
the mass-metallicity relation required
for the reproduction of the CM relation is  found to be less  promising
without appealing the less realistic assumption that $C_{\rm SF} \propto$
$L^{1.69}$ (Bekki \& Shioya 1997c).
This result implies either that other physical processes such as
the thermal and dynamical effects of type II supernovae
on the galactic dynamics
should be further
incorporated into the present merger model, or that the observed 
CM relation should not be interpreted so simply in terms
of mass-metallicity relation.
Thus it is one of our further study
to  investigate to what degree the dynamical and thermal effects
of supernovae-driven energy associated with the burst of
star formation during galaxy merging can modify the present results.
We should also reexamine the validity of the assumption
that the observed CM relation is actually mass-metallicity relation
of elliptical galaxy, as is described below.

 A growing number of observational and theoretical
studies aiming at breaking
the `age-metallicity degeneracy' suggest that the CM relation
of elliptical galaxies does not necessary mean the mass-metallicity relation.
Faber et al. (1995) suggest that even if the mean stellar metallicity
between elliptical galaxies are the same with each other, the  CM relation
can be equally reproduced with the assumption that
 the effective age of stellar populations is
progressively younger for less luminous ellipticals (`age effect').
Furthermore, Worthey et al. (1996) demonstrate that even if the typical 
or luminosity-weighted age of galaxy  
is  considerably spread (more than several Gyr) between  elliptical galaxies 
for a given luminosity,
the tightness of the  color-magnitude relation can be maintained 
with a specific relation that younger elliptical galaxies should be
more metal-enriched (the Worthey's 3/2 law).
This result implies that single burst model of elliptical galaxy formation
such as the galactic wind one, in which star formation
is assumed to be truncated at most within 2 Gyr,  can not be accepted at all
for a realistic model of elliptical galaxy formation.
Kodama \& Arimoto (1996) construct a specific
 model designed to  mimic the `age effect'
in order to  check the validity of the above 
suggestion and demonstrate that the pure `age effect' is not convincing
for reproducing both the tightness of the  color-magnitude relation 
observed
in the present cluster ellipticals and that  in the intermediate redshift cluster 
ellipticals. 
It is safe to say here that neither the pure `age effect' nor the
pure `metallicity  effect' (i.e., mass-metallicity relation)
could  explain a growing number of evidences 
suggesting the diversity of star formation history in elliptical galaxies.
We must wait more extensive observational studies which can confirm
whether the suspected age spread and the Worthey's 3/2 law,
both of which shed  new insight on the origin of the CM relation,
are robust and
does not depend on the galactic mass, environment, and redshift. 
We can say, however, that both the  age spread and the Worthey's 3/2 law 
could be  naturally explained by the present merger model
in a rather qualitative manner.
As is demonstrated in the previous sections, the epoch of galaxy merging
can affect both the mean age and metallicity of merger remnants in the sence
that later mergers are more likely to become ellipticals
with more metal-enriched and younger stellar populations.
This result suggests that the Worthey's 3/2 law can  be essentially ascribed to
the difference in the epoch of galaxy merging between galaxy mergers.
Furthermore, as is predicted by the semi-analytical models of
galaxy formation based upon the hierarchical clustering scenario
with CDM cosmogony (e.g., Baugh et al. 1996),
it is quite reasonable that the epoch of galaxy merging
differs from galaxies to galaxies.
Accordingly the luminosity weighted ages,
which might be dependent on the strength of the secondary starbursts in
galaxy merging, can be  naturally spread between elliptical galaxies formed by galaxy 
merging. 
Thus galaxy merging between late-type spirals,
which might be occurred in variedly different epoch,  is a promising
candidate that can explain the apparent diversity of star formation history
observed in real elliptical galaxy and nevertheless can
explain the tightness of the CM relation
of elliptical galaxies.

 
\section{Conclusion}

Main results obtained in the present  study are summarized as follows.

(1) Galaxy mergers with more rapid star formation  become ellipticals with
larger mean stellar metallicity, primarily because in
the mergers with more rapid gas consumption, a smaller  amount of metal-enriched 
gas is tidally stripped away during merging and consequently a 
larger  amount of the gas
can be converted into stellar component.  
This result is demonstrated not to depend so strongly on the  other parameters
such as the orbit configuration of galaxy merging and multiplicity of the mergers. 
These results suggest that the origin of the color-magnitude relation
of elliptical galaxies can be  closely associated with the details of
merging dynamics which depends on 
the rapidity of star formation
(thus on the galactic luminosity) in  galaxy mergers.

(2) Negative metallicity gradient fitted reasonably well by power-low can 
be reproduced by  dissipative galaxy mergers with star formation, 
which is in good  agreement with the recent observational  results
of elliptical galaxies.
The absolute magnitude of metallicity gradient in each merger remnant
depends  on the orbit 
configuration of each galaxy merging, 
suggesting that the observed dispersion in
 the absolute magnitude of
metallicity gradient for a given luminosity range of elliptical galaxies reflects
the diversity in the orbit configuration  of galaxy merging.
 
(3) Absolute magnitude of metallicity gradient  correlates  with that of 
age gradient in a merger
in the sence that a merger remnant with steeper negative metallicity gradient 
is more likely to show
steeper age gradient.
This result reflects the fact  that the degree of violent relaxation 
and gaseous dissipation during merging strongly affect both the  age
gradient and
metallicity one. 

(4) The outer part of stellar populations is both older and less  metal-enriched
than nuclei in an elliptical galaxy formed by galaxy merging with
less rapid star formation. 
Moreover
galaxy mergers with less rapid star formation are more likely to
become ellipticals with metal-poor gaseous halo.
This result suggests that the formation
of metal-poor $X$-ray halo actually observed in elliptical galaxies
can be essentially ascribed to the dissipative galaxy merging 
between late-type
spirals, and furthermore 
provides a clue to a solution for the iron abundance discrepancy problem
in elliptical  galaxies.

(5) The epoch of galaxy merging affects both the mean stellar metallicity 
and the mean stellar age in merger remnants:
Later galaxy mergers become ellipticals with both younger and more metal-enriched
stellar populations.
This result suggests that the origin of Worthey's 3/2 rule (Worthey et al. 1996), 
which is invoked in maintaining the tightness of the color-magnitude relation
of elliptical galaxies, can be understood in terms of the difference in the
epoch of galaxy formation and transformation, that is, the epoch of galaxy merging,
between elliptical galaxies.

(6) Luminosity dependence of chemical, photometric, and spectroscopic properties
in merger remnants, which is derived by adopting a specific assumption
on the luminosity dependence of the rapidity of star  formation of
galaxy mergers, does not match so reasonable well with that observed in
real elliptical galaxies.
This result implies that other fundamental physical processes expected to be
dependent on the galactic luminosity should be incorporated into the present
merger model for more successful comparison with observational trends
of luminosity-dependent chemical, photometric, and spectroscopic properties
of elliptical galaxies.

(7) As is described in the above (1) - (6),
the details of gas dynamics of galaxy merging, 
in particular, the tidal stripping of 
metal-enriched interstellar gas and the degree of gaseous dissipation during merging,
both of which depend on the star formation history of galaxy mergers,
are demonstrated to determine even the chemical and photometric
properties of merger remnants.
These results can not be obtained until both the chemical and dynamical evolution
during galaxy merging are solved numerically in a reasonably self-consistent way.

\acknowledgments

We are grateful to the referee for valuable comments, which contribute to improve
the present paper.
K.B. thanks to the Japan Society for Promotion of Science (JSPS) 
Research Fellowships for Young Scientist.

\newpage

\appendix
\section{The expected luminosity dependence  of the parameter $C_{\rm SF}$}
\subsection{The importance  of  $C_{\rm SF}$}
In the present study, we adopt the assumption
that  more massive (luminous)  elliptical galaxies are formed by galaxy $major$ merging 
between more massive (luminous) late-type spirals, and thereby investigate
how the difference of galactic mass (luminosity) of progenitor disks in a merger
can affect the chemical and photometric properties of merger remnants. 
It is reasonable and realistic 
that if the mass of merger progenitor is different  between 
galaxy mergers, there could be remarkable differences between mergers in 
the fundamental physical processes  related to the chemical 
evolution of galaxies, such as the duration, strength, time-scale
of star formation (star formation history), dissipative dynamics of interstellar
gas, thermal and dynamical effects
of accumulated thermal energy driven supernovae events,
and the chemical mixing driven by the dynamics of galaxy merging.
Accordingly, we should investigate such physical processes expected to
be dependent on galactic mass (luminosity) and thereby clarify the relative
importance of these processes.
Among these, what has been the most extensively examined in the previous studies is
the thermal and dynamical `feedback' effects associated with the supernovae
events, primarily because the ratio of the accumulated
thermal energy driven by the supernovae
to the total potential energy of galaxies has been 
considered to depend predominantly on  galactic luminosity.
Although there are a number of important issues which should be addressed extensively,
we here focus on  the difference in the star formation history
between galaxy mergers with different  mass (or luminosity), 
and thus other important issues will be explored in our future  papers. 

 The adopted `working hypothesis' that  galactic mass  predominantly
determines the star formation history of elliptical galaxies is quite realistic and
reasonable, since a growing number of observational studies support 
 this hypothesis.
For example, the line ratio of [Mg/Fe], which can be interpreted as 
 the strength of the past  activity of type-II supernovae relative to that of
the type I supernovae,  implies that  more luminous elliptical galaxies
are more likely to truncate their star formation earlier (e.g., Worthey et al. 1992).
Furthermore, as is suggested by the analysis of H$\beta$ line index, 
the epoch, the strength, and the duration of the latest star formation
of elliptical galaxies seem quite diverse,
implying that the classical single burst picture  of elliptical galaxy formation
becomes less attractive (Faber et al. 1995).
It should be also emphasized that the $B-H$ color in more
luminous disk galaxies,
which are considered to be merger precursors of more luminous ellipticals
in the present study,
is found to be redder than the less luminous one (Wyse 1982; Bothun et al.
1985; Gavazzi \& Scodeggio 1996).
These lines of observational evidences  strongly motivate us to clarify
the important roles of star formation history in determining the chemical
and photometric properties as well as dynamical and kinematical ones
in elliptical galaxies.

 In the present paper, we focus particularly on the time-scale of gas consumption
by star formation $relative$ $to$ $the$ $dynamical$ $time-scale$ of 
galaxy mergers: 
It should be emphasized here that  
the time-scale of gas consumption by star formation is not  considered to
be an important determinant but the time-scale of gas consumption
by star formation $relative$ $to$ $the$ $dynamical$ $time$-$scale$ is to be 
an important determinant.
The reason for this consideration is explained as follows.
The time-scale of gas consumption by star formation is the typical
time-scale within which heavy elements are produced  by star formation
and mixed into interstellar medium.
The dynamical time-scale is the typical time-scale  within which
violent relaxation during galaxy merging cause  the  
mass and angular momentum redistribution and thus the 
dynamical mixing of heavy elements produced by star formation.
Therefore, if the dynamical time-scale is much larger than 
the gas consumption  time-scale,  star formation can proceed quite efficiently
before the system reach the dynamical equilibrium,  and consequently the
larger amount of heavy elements produced by the star formation
can suffer  more effective dynamical mixing of the heavy elements
during galaxy merging. 
Thus, since the dynamics of galaxy merging can strongly affect even the chemical
evolution in the present merger model, the ratio of the above two time-scales
are expected to be more essential for the chemodynamical evolution of
galaxy mergers.
For convenient, the inverse ratio of the time-scale of gas consumption
by star formation to the dynamical time-scale is referred to as the rapidity
of star formation and represented by the parameter  $C_{\rm SF}$.
It is the $C_{\rm SF}$ that we consider to depend strongly on the galactic
luminosity, accordingly, we investigate the most extensively the important
roles of  $C_{\rm SF}$ in determining fundamental characteristics of
elliptical galaxies in the present study.

\subsection{A possible luminosity dependence of $C_{\rm SF}$}

The expected luminosity (mass) dependence of  $C_{\rm SF}$ is described 
as follows.
The parameter $C_{\rm SF}$ is set to be proportional to 
 $T_{\rm dyn}$/$T_{\rm SF}$,
where  $T_{\rm dyn}$ and $T_{\rm SF}$  are the  dynamical time-scale
and the time-scale of gas consumption by star formation,
respectively.
We here define  the mass of a galactic disk, the total mass   of 
luminous  and dark matter,  and size of the progenitor as 
$M_d$, $M_t$, and $R_d$, respectively.
We consider here 
that gas mass in a disk is equal to $M_d$ for simplicity. 
The $T_{\rm dyn}$ is given as  
\begin{equation}
T_{\rm dyn} \propto R_d^{3/2} M_t^{-1/2}
\end{equation}
Provided that the coefficient in the Schmidt law is not dependent on the
galactic mass (or luminosity), we can derive $T_{\rm SF}$ as follows. 
\begin{equation}
T_{\rm SF} \propto \Sigma^{1 - \gamma} \; ,
\end{equation}
where $\Sigma$ is the surface density of the gas disk. 
The parameter $\gamma$ is the exponent of Schmidt law, which is the same as 
that used in previous subsections. 
Assuming the Freeman's law and the constant ratio of $R_d$ to 
the scale length of exponential disk, 
we derive
\begin{equation}
\Sigma \propto M_d R_d^{-2} \sim {\rm const.} 
\end{equation}
Assuming that the degree of self-gravity of a galactic disk is described as 
\begin{equation}
M_t \propto M_d^{(1 - \beta)} \; \; ,
\end{equation} 
then we can derive 
\begin{equation}
C_{\rm SF} \propto M_d^{1/4 + \beta/2} \; \; .
\end{equation}
Since $\beta$ is considered to have positive value 
($\beta = 0.6$: Saglia 1996), 
this relationship predicts that $C_{\rm SF}$ becomes larger as 
$M_d$ increases.
Furthermore, if we  
assume the constant mass to light ratio for luminous matter,
we can  obtain the luminosity dependence of $C_{\rm SF}$.
In this case, the dependence is described as  $C_{\rm SF}  \propto L_d^{0.55}$,
where $ L_d^{0.55}$ is disk luminosity, for
$\beta = 0.6$.
Alternatively,
if we adopt the observed trend that more luminous disks have large surface 
density, such as $\Sigma \propto M_d$ (McGaugh \& Blok 1997), we can obtain
$C_{\rm SF} \propto M_d^{1/2 + \beta/2}$.
Thus, these simple theoretical arguments   suggest that larger (or more luminous)
disk galaxies are more likely to have larger values of $C_{\rm SF}$.

\placefigure{fig-18}
\placefigure{fig-19}
\placefigure{fig-20}

\section{Structure  and morphology of merger remnants}
Although our main purpose of the present paper is not concerned
with the dynamical properties of elliptical galaxies,
we briefly describe  how successfully the present merger model
can reproduce the luminosity-dependent dynamical properties of
elliptical galaxies. 
The reason for this is that
since in the present merger model,
the details of chemical evolution are strongly affected by the dynamical
evolution of galaxy mergers,  
observing the dynamical evolution of dissipative galaxy merging
with star formation would help us to understand more clearly the present results
on chemical evolution of galaxy mergers.

\subsection{Dependence on $C_{\rm SF}$}
First we present  morphological, structural,   
and kinematical properties of merger remnants and their
dependence on $C_{\rm SF}$.
As is described in details
by Bekki \& Shioya (1997a, b), the galaxy mergers
with larger $C_{\rm SF}$
are found to be more likely to become elliptical galaxies with less strongly supported by 
global rotation, less strongly self-gravitating, smaller central surface
brightness, larger cores, and boxy isophotal shape. 
Figure 18 describes the morphological evolution of the model with $C_{\rm SF}$ = 0.35
(Model B1).
As the dynamical interaction between star-forming disk galaxies becomes
stronger, an appreciably amount of metal enriched gas  is 
 tidally stripped away from the disks and finally surrounds the 
developed elliptical galaxy without further star formation and chemical enrichment.
The developed  metal-poor gaseous halo might be observed as the hot and metal-poor X-ray
halo of elliptical galaxies, which has been recently revealed by the 
$ASCA$
 (e.g., Matsumoto et al. 1997). 
The final morphology of merger remnants depends strongly on the rapidity of
star formation represented by the $C_{\rm SF}$,  as is shown in Figure 19:
A galaxy merger with less rapid star formation becomes a more compact 
elliptical galaxy. 
The density profiles of merger remnants are well fitted by $R^{1/4}$ law,
however, the systematical deviation from the $R^{1/4}$ law  is also observed 
depending on values of $C_{\rm SF}$,
as is shown in Figure 20. 
Unlike the late galaxy merger in which the gas mass fraction of
the precursor disk is less than  0.2, the present merger model with highly
dissipative nature of interstellar gas
succeeds 
in boosting up more significantly the central density of merger remnant  
compared with isolated disk models (Model A1 and A2).
This result
suggests the problems related to phase space density of elliptical galaxies
(Ostriker 1980; Carlberg 1986) can be successfully
 resolved in high-redshift galaxy mergers,
as has been already  indicated by  Kormendy \& Sanders (1992).
These results imply that galaxy mergers with star formation and gaseous dissipation,
which might be occurred the most frequently at higher redshift,
can naturally explain the fundamental dynamical and kinematical properties
of elliptical galaxies (For further details, see Bekki \& Shioya (1997a, b).).

\subsection{Dependence on galactic luminosity}
 By adopting a specific assumption on the luminosity dependence of the parameter
$C_{\rm SF}$, we can observe how successfully the present merger
can reproduce the $luminosity-dependent$ morphological,
structural, and kinematical properties of elliptical galaxies.
Although describing  the dynamical properties of elliptical galaxies 
and their dependence on galactic luminosity is
not main purpose of this paper, however, we beforehand and briefly discuss them,
since we  will point out  advantages and disadvantages of the 
present merger model in reproducing $both$  dynamical properties 
and chemical ones of elliptical galaxies in  the section \S 4. 
Observational results indicate that  more luminous elliptical galaxies
are more likely to be less rotationally supported (Davies et al. 1983), 
possess less luminous
surface densify (Djorgovski et al. 1996) and smaller
phase space density (Carlberg 1986), have boxy isophote (Kormendy \& Bender 1996)
and larger cores (Kormendy \& Djorgovski 1989).
If we assume that $C_{\rm SF} \propto L^{0.55}$, that is, 
more luminous elliptical galaxies are formed by galaxy mergers with
more rapid star formation, 
our numerical results are consistent  with the above observed trends of
$luminosity-dependent$ dynamical properties of elliptical galaxies at least
in a qualitative manner.
For example, the observed trend that the galactic core radius is more
likely to be larger
for the more luminous elliptical galaxy can be naturally explained
by the present merger model that more luminous ellipticals are
formed by galaxy $major$ merging between more luminous spirals. 
The first reason for this
is that since  more luminous spirals $initially$ are more likely
to have
larger `cores' (or smaller central phase space density) 
as is implied by the
Freeman's law and the relation between galactic luminosity and scale-length
(e.g., McGaugh and Blok 1997),
galaxy mergers between spirals with larger cores are more likely to become
ellipticals with larger cores (or smaller central phase space density). 
The  second reason is that 
 more luminous
galaxy mergers are less dissipative owing to the more rapid consumption
of interstellar gas, and consequently  become ellipticals with less degree of
central concentration.
Hence the luminosity-dependent central structure in elliptical galaxies
could  reflect the luminosity-dependent dynamical structure of disk galaxies
and the star formation history of galaxy mergers.
Our numerical results 
imply furthermore that any particular physical processes such as the dynamical
heating of the central cores by 
binary black holes (Ebisuzaki,  Makino, \& Okumura  1991) 
are  not  necessarily required  for the formation
of larger cores in more luminous elliptical galaxies. 
Thus, our numerical results 
strongly suggest that the
luminosity-dependent dynamical structure of elliptical galaxies
can be understood in the context of the difference in
luminosity-dependent dynamical structure of progenitor disks
and the star formation history between
gas-rich galaxy mergers. 
This furthermore indicates that
total amount of gaseous dissipation and the degree of violent relaxation
during merging, both of which are determined basically by the star formation
history of galaxy mergers,
are important factors which can affect the luminosity-dependent morphological,
structural, and kinematical properties of elliptical galaxies.
The validity of the suggested point of view that luminosity-dependent dynamical
structures of elliptical galaxies are essentially ascribed to the
difference in the  star formation history between elliptical
galaxies formed by dissipative
galaxy merging has been  also investigated  by (Bekki \& Shioya 1997b)
 in the context of the origin
of the fundamental plane of elliptical galaxies. 
Although more elaborated numerical studies including more realistic
implementation of star formation and gaseous dissipation are definitely
required for confirming the validity of the  results obtained in
the present study,
the merger scenario that more luminous ellipticals
are formed by dissipative and $major$ galaxy merging between
more luminous spirals seems likely in the context of
the reproduction of the observed luminosity-dependent dynamical
properties of elliptical galaxies.

\subsection{Self-consistency of the present merger model}

 Our final goal is to construct a more realistic model
which can clearly explain both the structural and kinematical properties,
such as the origin of the fundamental plane (Djorgovski \& Davis 1987;
Dressler et al. 1987), boxy-disky dichotomy (Kormendy \& Bender 1996),
and surface brightness - effective radius relation (Djorgovski et al. 1996),
and the chemical and
photometric properties, such as the color-magnitude relation (Faber 1973;
Visvanathan \& Sandage 1977) and
luminosity-dependent radial gradient of metallicity, age, and color
 (Peletier et al. 1990;
Davies et al. 1991).
Therefore it is essential to discuss how successfully
the present merger model
has actually reproduced  or would reproduce
both the dynamical and kinematical properties
and chemical and photometric ones in a self-consistent way.
What is the most vital in addressing this crucial issue is to 
try to explain  both the $luminosity-dependent$ dynamical and chemical properties
of elliptical galaxies.
Since the present study is the first step toward the  complete understanding
of elliptical galaxy formation, 
we first make only a qualitative comparison of the present results with observational
one, then we point out the advantages and disadvantages of the present merger model
in reproducing both the $luminosity-dependent$
dynamical and chemical properties of real elliptical galaxies. 
In the following discussions, we adopt the assumption  that  galaxy mergers with larger 
$C_{\rm SF}$ become more luminous  elliptical galaxies,
mainly because this assumption is  realistic and  
essential for explaining the morphological and kinematical properties
of elliptical galaxies formed by galaxy merging (Bekki \& Shioya 1997a, b).

As is described in details by Bekki \& Shioya (1997a, b),
more luminous elliptical galaxies formed by dissipative galaxy merging
with star formation are less rotationally supported, less compact,
less strongly self-gravitating, and  more likely to have
boxy isophotal shape.
In the present study, more luminous elliptical galaxies are found to be more likely to
be redder in global color than less luminous ones.
These obtained results concerning the luminosity
dependence of  structural, dynamical and photometric properties
agree at least qualitatively  with the observed trends of 
elliptical galaxies in a self-consistent manner.
These results strongly suggest that both the  origin of 
luminosity-dependent dynamical and morphological structure and that of
the luminosity-dependent chemical,  photometric,
and spectroscopic  properties in elliptical galaxies 
are closely associated with the star formation history, in particular,
with the rapidity of star formation in galaxy mergers.
Furthermore these results imply that dissipative galaxy merging with star formation,
which might be the most frequently occurred in higher redshift, 
has  a  number of advantages in reproducing
the nature of elliptical galaxies  than we have previously expected.

 What we should note here is that the present merger model also has a number of
problems concerning  the self-consistent
reproduction of structural and chemical properties of
elliptical galaxies.
One of the problems that should be foremost resolved is on the luminosity
dependence of the central
surface brightness (or density)
and that of the central stellar metallicity in elliptical galaxies. 
Observational studies have revealed that  more luminous elliptical galaxies
show both less luminous central (or effective) surface brightness, as  is indicated
by the Kormendy relation (e.g., Djorgovski et al. 1996), and
the larger central stellar metallicity, 
as is indicated by $\rm Mg_{2}$ - $\sigma$ relation (Burstein et al. 1988).
This fundamental tendency of elliptical galaxies has not yet be reproduced by the present
merger model, 
because the present model predicts that  more luminous elliptical galaxies
show less luminous central surface (and larger cores)
brightness but the smaller central stellar
metallicity compared with  less luminous ones.
This sort of failure can be seen also in the dissipative collapse models of Larson (1975)
and Carlberg (1984), in which elliptical galaxies with larger central stellar metallicity
show more luminous central surface brightness (or smaller cores).
We consider 
these apparent failures are probably due to the ill approximation of the adopted
instantaneous recycling, in which chemical enrichment is assumed to proceed
considerably faster than the dynamical evolution.
Actually the instantaneous recycling approximation is expected to give
an undesirable result to  less luminous elliptical galaxies,
in which  the  dynamical time-scale of the systems is comparable
or only slightly larger than the typical life-time of massive stars 
(a few $10^{7}$ yr). 
Specifically, for less luminous galaxies, dynamical evolution such as the
violent relaxation and redistribution of angular momentum of gas and stars,
which plays a vital role in transferring and mixing
dynamically  chemical components,
probably finishes earlier before the onset of efficient chemical 
enrichment driven by type II supernovae.
As a result of this, the  dynamical transfer of chemically enriched components
into the inner region of galaxies,
which determines the magnitude of the central stellar metallicity,
are less likely for less luminous elliptical
galaxies.
This indicates the ratio of the typical time-scale of 
chemical enrichment (a few $10^{7}$ yr) to the dynamical time-scale
and its luminosity dependence are more essential for the 
chemodynamical evolution of less
luminous elliptical galaxies. 
This kind of chemical evolution expected for less luminous ellipticals (mergers)
is probably not modeled by the present chemodynamical model in a  proper way,
accordingly the present model with instantaneous recycling
should be greatly modified to a more realistic one.
Hence, it is our future and essential study to consider the importance of the
ratio of the time-scale of chemical enrichment  to the dynamical time-scale in
elliptical galaxy formation 
and then to investigate again whether or not the remaining problem 
of the present merger model can be resolved in the more sophisticated models
of chemodynamical evolution. 

Thus, the present numerical study suggests that
merger remnants
with larger central metallicity 
are more likely to show   larger 
central surface density, which  appears to disagree with the observational
trend of real elliptical galaxies.
This apparent failure of the present merger model 
accordingly   leads  us to  the  conclusion that other important 
physical process associated with dissipative galaxy merging, 
such as the thermal and dynamical feedback effects of star formation,
should be included for more successful comparison 
with the observational results.
The alternative conclusion is
that the instantaneous recycling approximation adopted in the present study
could not  be so appropriate for analyzing both chemical and dynamical evolution
in galaxy mergers, especially for
less luminous galaxy mergers in which typical time-scale of chemical 
enrichment  is
comparable or larger than the dynamical time-scale of the mergers.

\clearpage





\figcaption{Snapshots  of an isolated disk model  with 
$C_{\rm SF}$ = 0.35 (Model A1) projected onto $xy$ plane (upper)
and onto $xz$ plane (lower)  at $T = 15.0$ in our units
for stellar component (new stars).
Note that a stellar bar forms in the central part of the disk.
\label{fig-1}}

\figcaption{Time evolution of star formation rate in units of $M_{\rm d}/t_{\rm dyn}$
for merger model with $C_{\rm SF}$ = 0.35, Model B1 (open circles)
 and 3.5, Model  B5 (open triangles).
$M_{\rm d}$ and $t_{\rm dyn}$ denote the initial mass of a progenitor disk
and the dynamical time of the disk, respectively.
\label{fig-2}}

\figcaption{Distribution of the 
stellar metallicity ($Z_{\ast}$) and the epock of star formation
($T_{\ast}$)
for the model with $C_{\rm SF}$ = 0.35 (Model B1, upper) and
3.5 (Model B5, lower) at $T$ = 15.0
in our units. The vertical height of each line represents the total 
number of stellar particles  with a given  $Z_{\ast}$ and $T_{\ast}$. 
Younger stellar particles  have larger values of $T_{\ast}$. 
\label{fig-3}}

\figcaption{
Time evolution of mean stellar metallicity ($<Z_{\ast}>$)
for model with  $C_{\rm SF}$ = 0.35, Model B1 (open circles)
 and 3.5, Model B5 (open triangles).
\label{fig-4}}

\figcaption{Time evolution of gas mass fraction ($f_{g}$) 
for isolated disk models, Model A1 and A2 (solid lines, in upper
panel)
and for merger models, Model B1 and B5 (solid lines, in lower panel).
For comparison, the time evolution of $f_{g}$   
for isolated disk models without dynamical evolution, Model A3 and A4 
is presented by dotted lines in each panel. 
Open  circles and triangles  denote the model with $C_{\rm SF}$ = 0.35
and that with $C_{\rm SF}$ = 3.5, respectively.
It should be noted that although the final gas mas fraction at $T$ = 15.0
is not so different between the two
isolated disk models without dynamical evolution (Model A3 and A4),
it is remarkably different between isolated disk models (Model A1 and A2)
and between  merger models (Model B1 and B5).
\label{fig-5}}

\figcaption{
Distribution of stellar metallicity for models with  
$C_{\rm SF}$ = 0.35, Model B1, (upper) and 3.5, Model B5 (lower) at $T=15.0$ 
in our units (solid lines). For comparison,
the stellar metallicity distribution predicted from the Simple one-zone
model for each
model is also plotted by asterisks ($\ast$) 
in the same panel. 
\label{fig-6}}

\figcaption{
Radial gradient of stellar metallicity (solid line) and that of gaseous one (dotted
line)
for the model with  $C_{\rm SF}$ = 0.35, Model B1 (open circles), and 3.5,
Model B5 (open triangles)  at $T=15.0$ in our units.
\label{fig-7}}

\figcaption{
Radial gradient of the mean epoch of star formation ($<T_{\ast}>$) 
for the models with  $C_{\rm SF}$ = 0.35, Model B1 (open circles), 
 and $C_{\rm SF}$ = 3.5,
Model B5 (open triangles) at $T=15.0$.
This figure describes the age gradiant in each merger remant.
\label{fig-8}}

\figcaption{
The shapes of isophotes contours (solid line) and isochromes ones (dotted line)
projected onto $xy$ plane  for the central region
of merger remant (within $R \leq 1.0$ in our units)
at $T=15.0$ in Model B2.
\label{fig-9}}

\figcaption{
Dependence of final mean stellar metallicity (at  $T=15.0$   in our units)
on the $C_{\rm SF}$ for standard models, Model B1 $\sim$ B5 (filled circles),
for models
with different chemical mixing length,
Model B6 $\sim$  B9 (open circles),  for models with different
epock of galaxy merging, Model B10 $\sim$ B15 (open triangles),
for models with different orbit configuration, Model C1, C2, D1, D2, E1, 
E2, F1 and F2 (open squares), and for multiple mergers, Model G1 and
G2 (crosses).
\label{fig-10}}

\figcaption{The same as Figure 10 but for the dependence of mean epoch of
star formation ($<T_{\ast}>$).
\label{fig-11}}

\figcaption{The same as Figure 10 but for the dependence of  
radial gradient of stellar metallicity.
\label{fig-12}}

\figcaption{The same as Figure 10 but for the dependence of radial
gradient of mean star formation epoch ($<T_{\ast}>$).
\label{fig-13}}

\figcaption{Dependence of the mean stellar metallicity ($<Z_{\ast}>$)
on the mean epoch of star formation ($<T_{\ast}>$) in merger remants
for models with different merging epochs. 
Open circles and triangles represent the models with $C_{\rm SF}$ = 0.35
and 3.5, respectively.
\label{fig-14}}

\figcaption{The dependence of the radial gradient of the
mean epoch of star formation  ($<T_{\ast}>$) on that of  
stellar metallicity  ($<Z_{\ast}>$) for 
all merger models. The notation of each mark is exactly the same as that
presented in the Figure 10.
\label{fig-15}}

\figcaption{
Radial distribution of the integrated color, $U-R$ (solid line) 
and $B-R$ (dotted) at 13 Gyr 
for the fiducial model with  $C_{\rm SF}=1.0$. 
\label{fig-16}}

\figcaption{
Color-magnitude relation for all  merger models 
at 13 Gyr.
The notation of each mark is exactly the same as that presented in the Figure 10.
For comparison, the integrated global color expected from the Simple
one-zone model
is also plotted by asterisks for five standard  models with different $C_{\rm SF}$
(Model B1 $\sim$ B5) in the same figure. 
A solid line represents the observed color-magnitude relation derived
from Bower et al. (1992).
\label{fig-17}}

\figcaption{
Morphological evolution of a galaxy merger with $C_{\rm SF}$ = 0.35 (Model B1)
projected onto $xy$ plane for halo (top), interstellar
gas (middle) and new stellar component
(bottom). $T$ indicates the time in our units. Note that an appreciable amount of 
interstellar gas 
is tidally stripped away during merging to form gaseous halo.
\label{fig-18}}

\figcaption{Snapshots  of stellar component in
merger remnants projected onto $xz$ plane
at $T = 20.0$ in our units
for model with $C_{\rm SF}$ = 0.35, Model B1  (upper),  and $C_{\rm SF}$ = 3.5,
Model B5  (lower).
\label{fig-19}}

\figcaption{Radial distribution of stellar component 
projected  onto $xy$ plane for merger remnants (solid lines) in the model with
$C_{\rm SF}$ = 0.35 (Model B1) and $C_{\rm SF}$ = 3.5 (Model B5)
and for  isolated disks (dotted ones)
with $C_{\rm SF}$ = 0.35 (Model A1) and $C_{\rm SF}$ = 3.5 (Model A2)
at $T$ =15.0 in our units.
Models with $C_{\rm SF}$ = 0.35 and those with $C_{\rm SF}$ = 3.5
are represented by open circles and by open triangles,
respectively. 
\label{fig-20}}

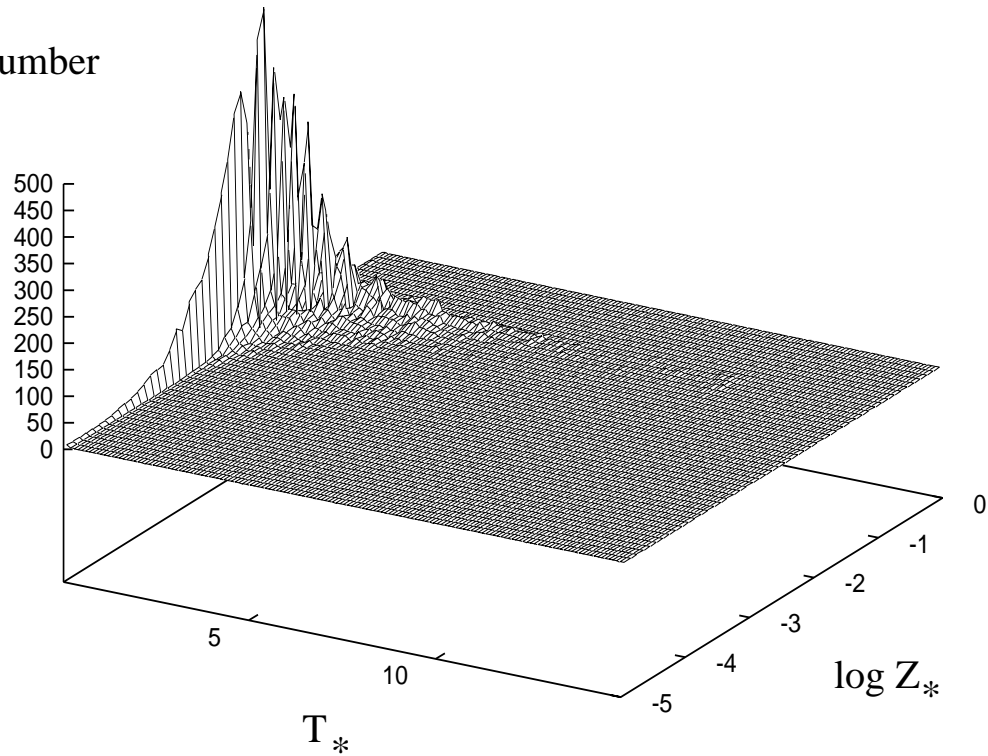
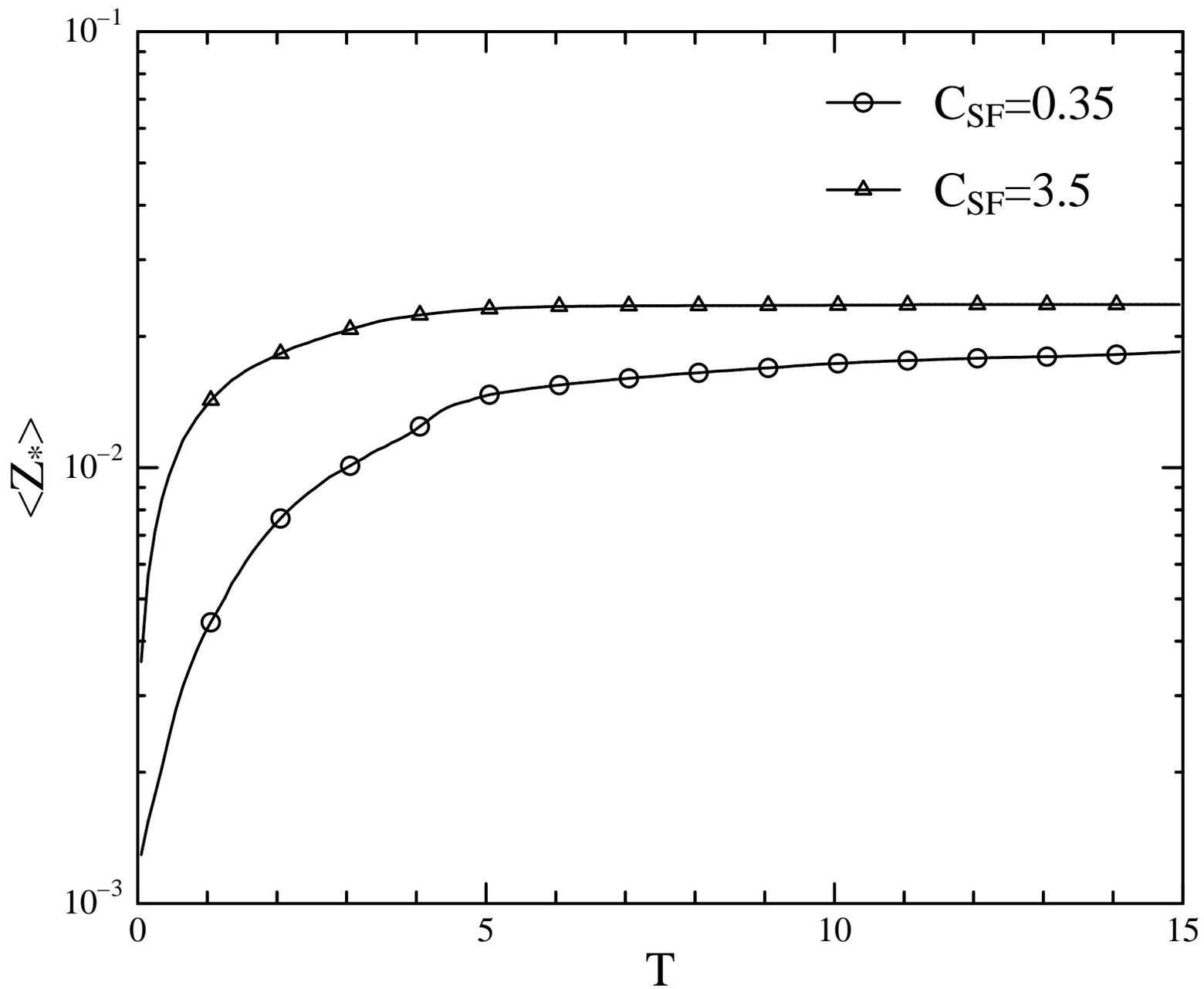
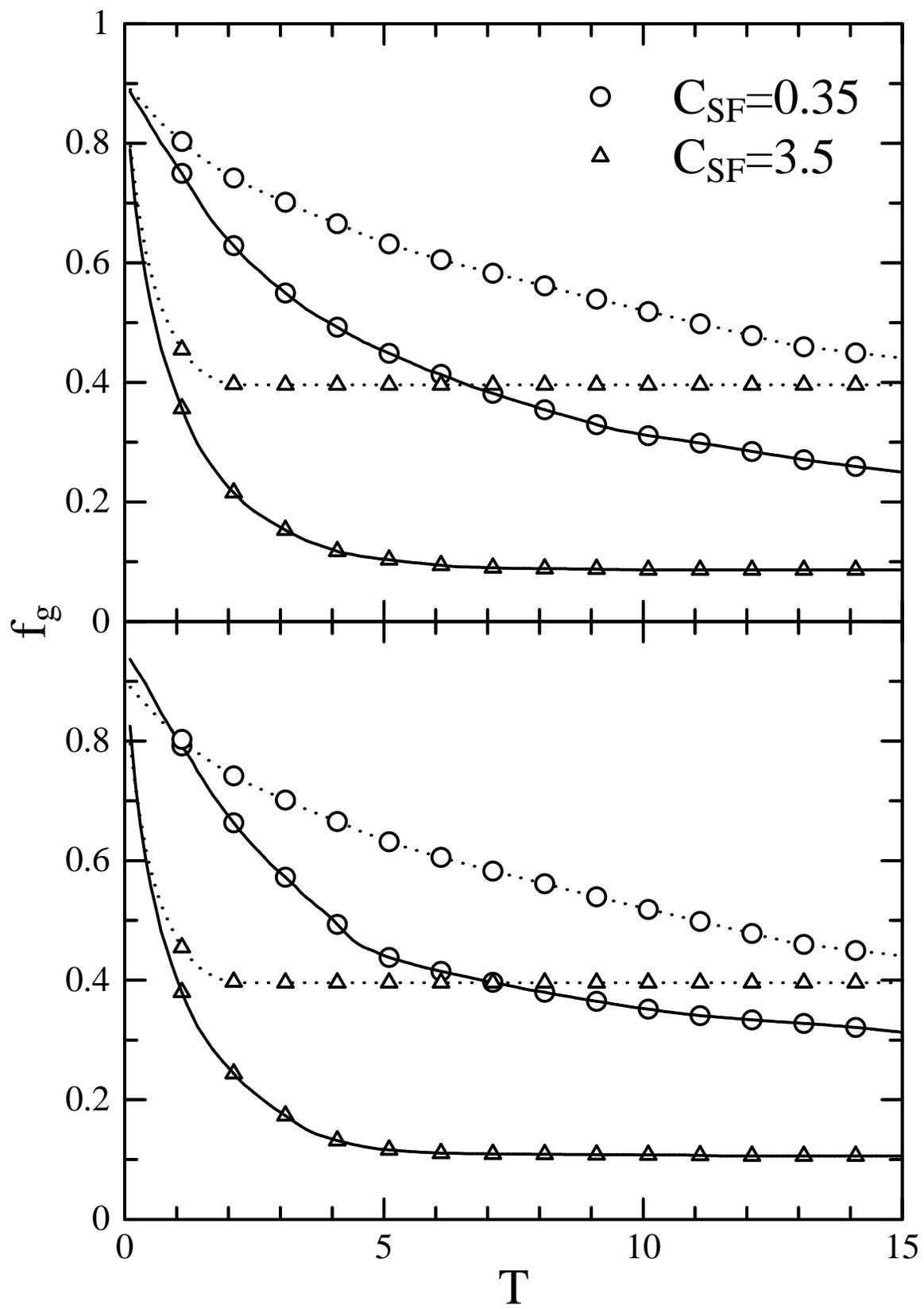
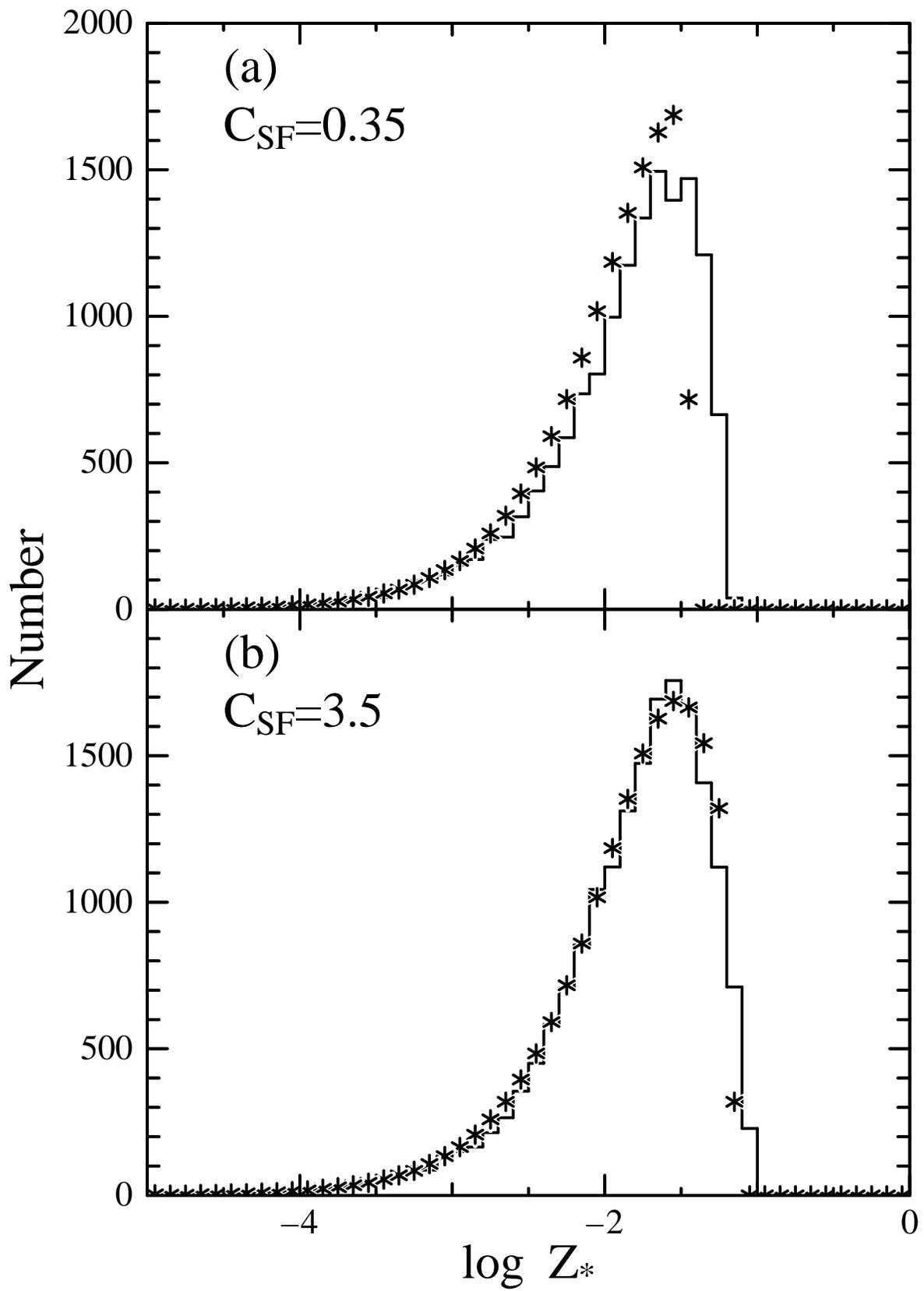
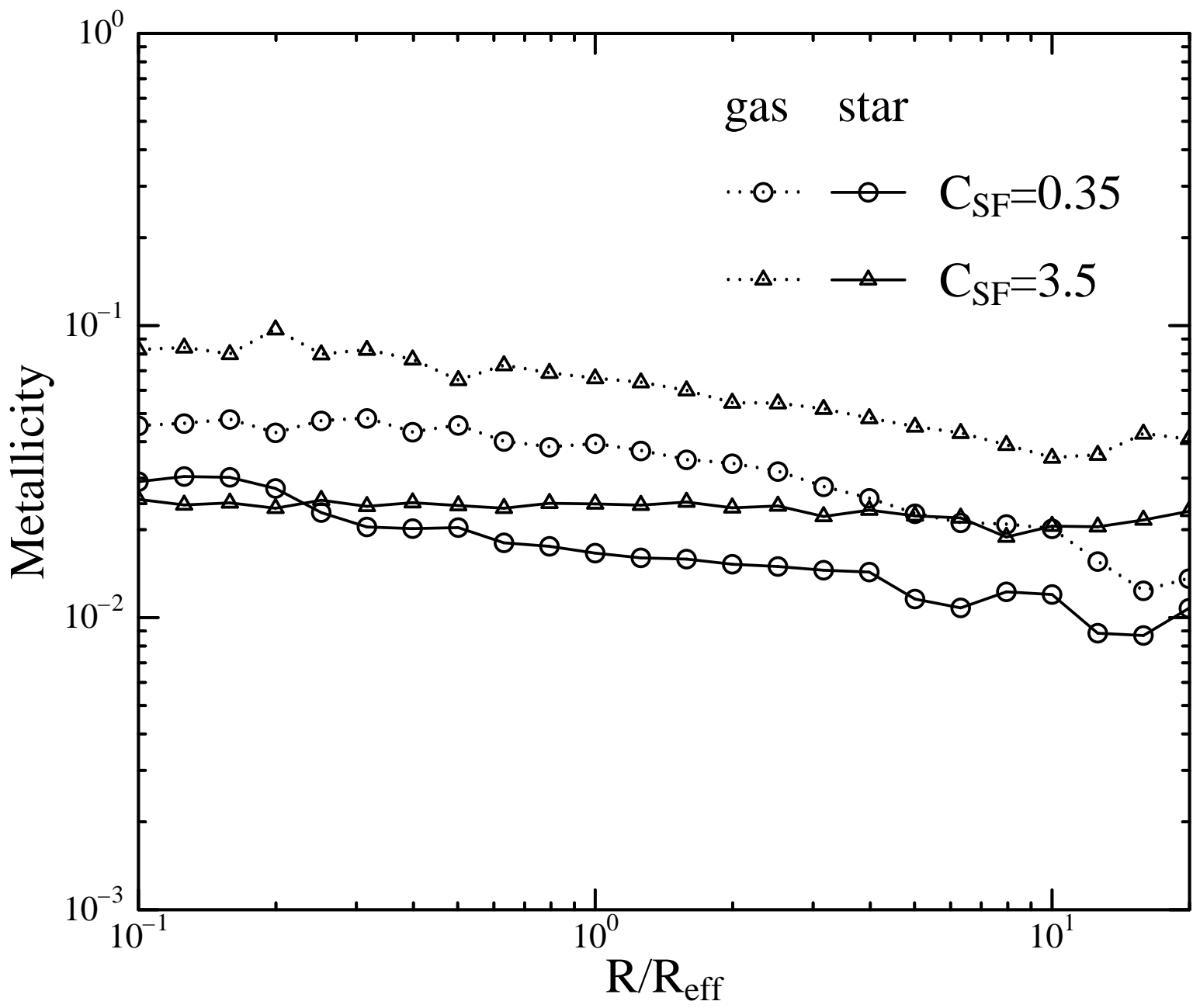
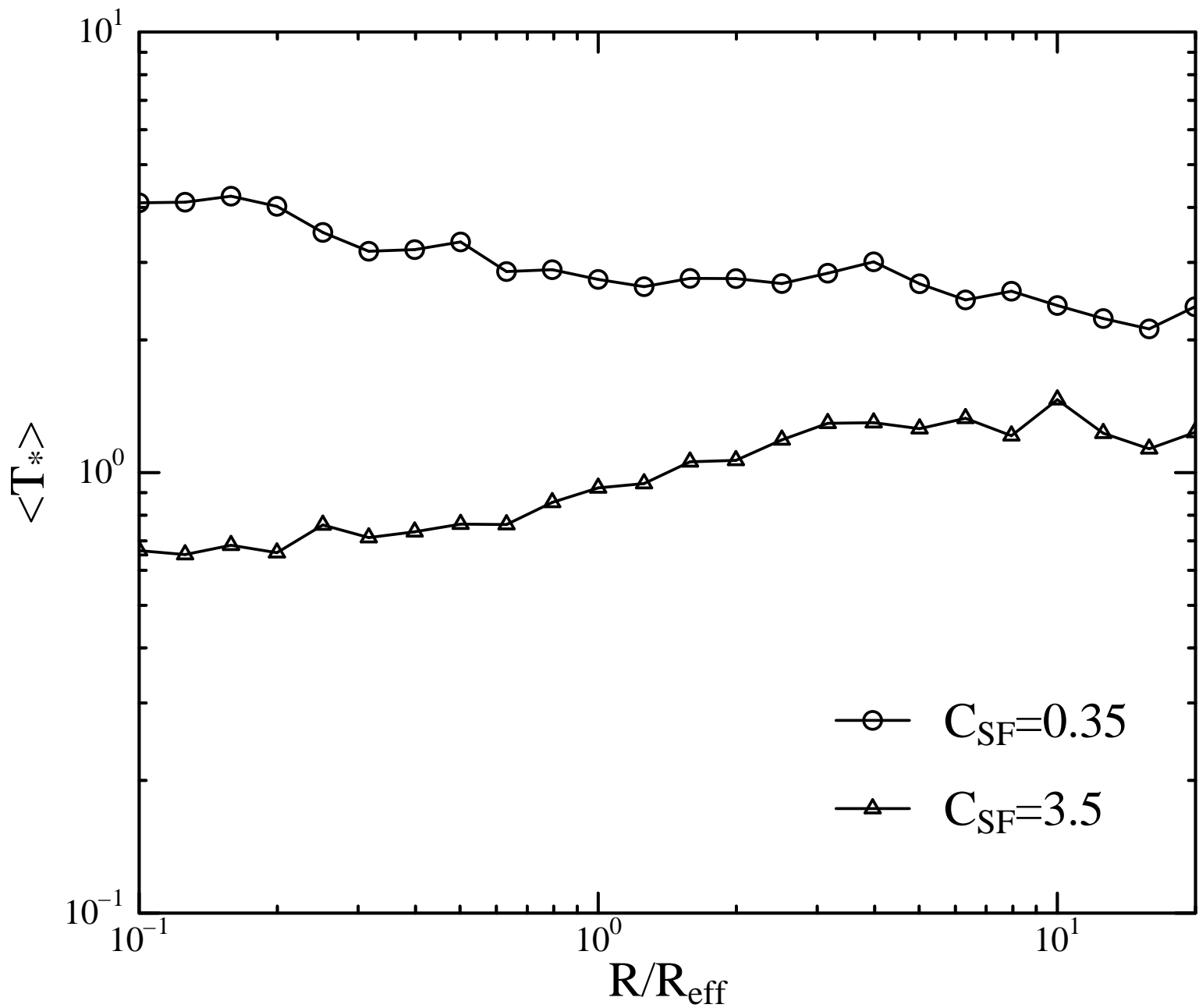
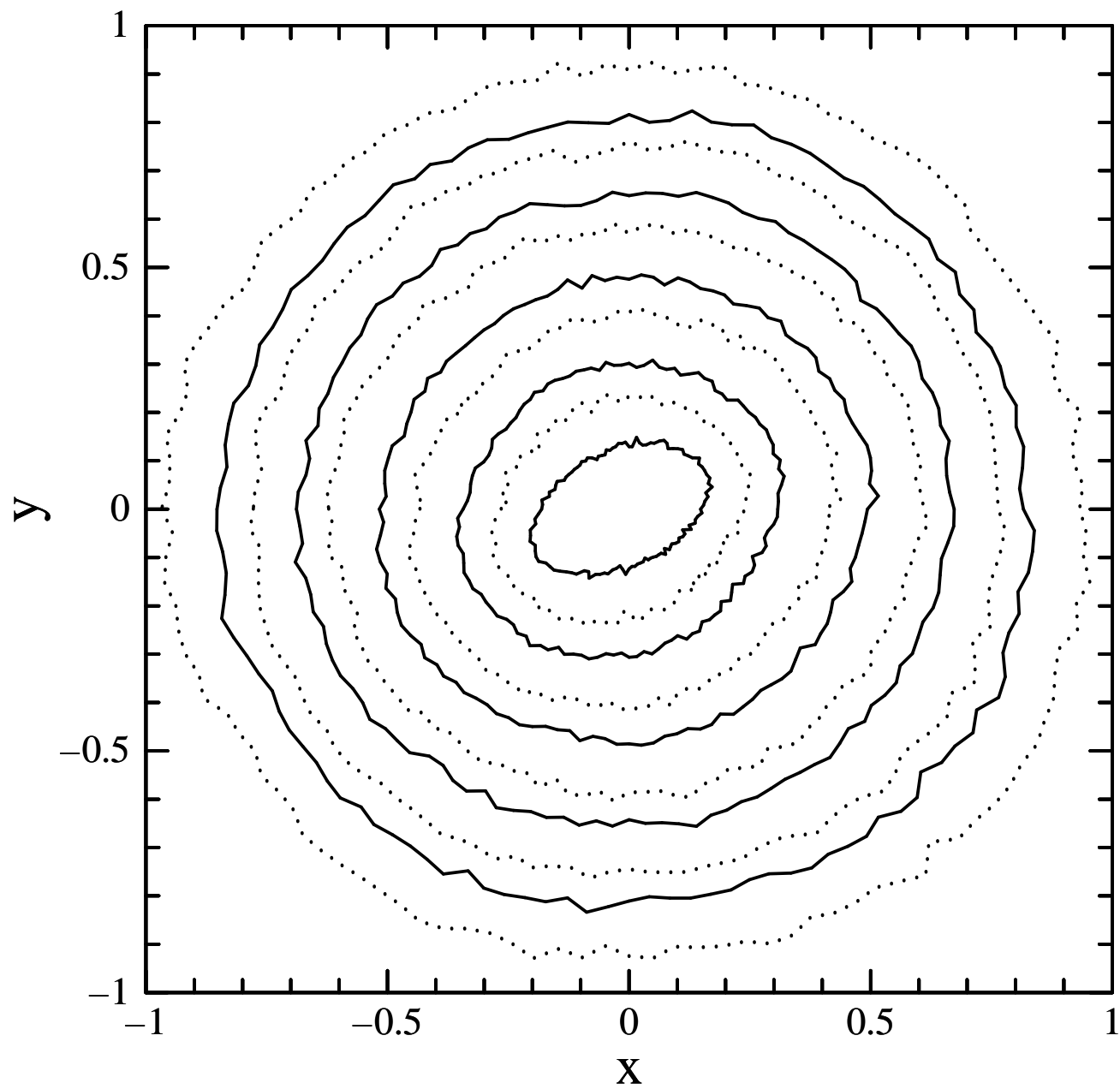
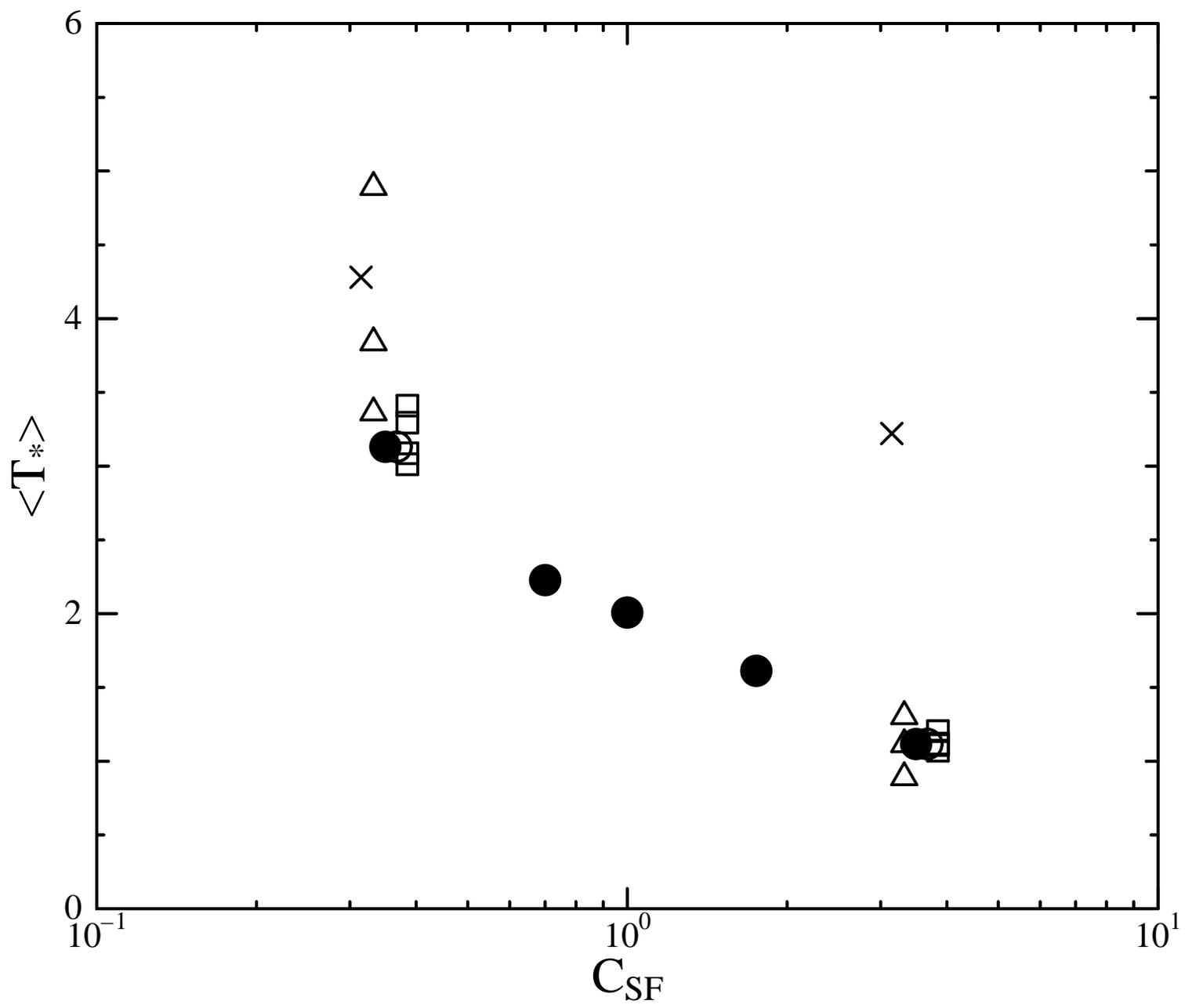
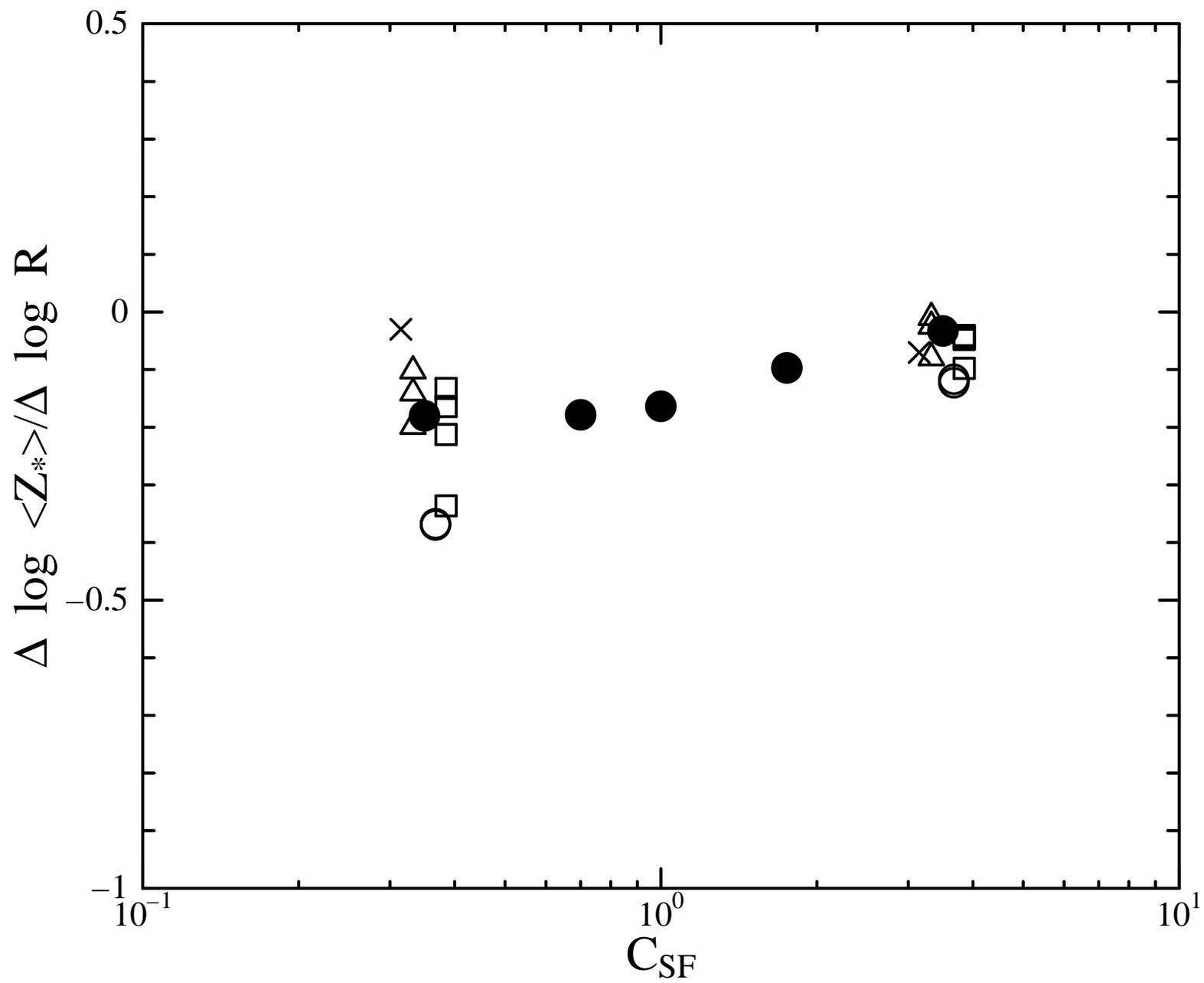
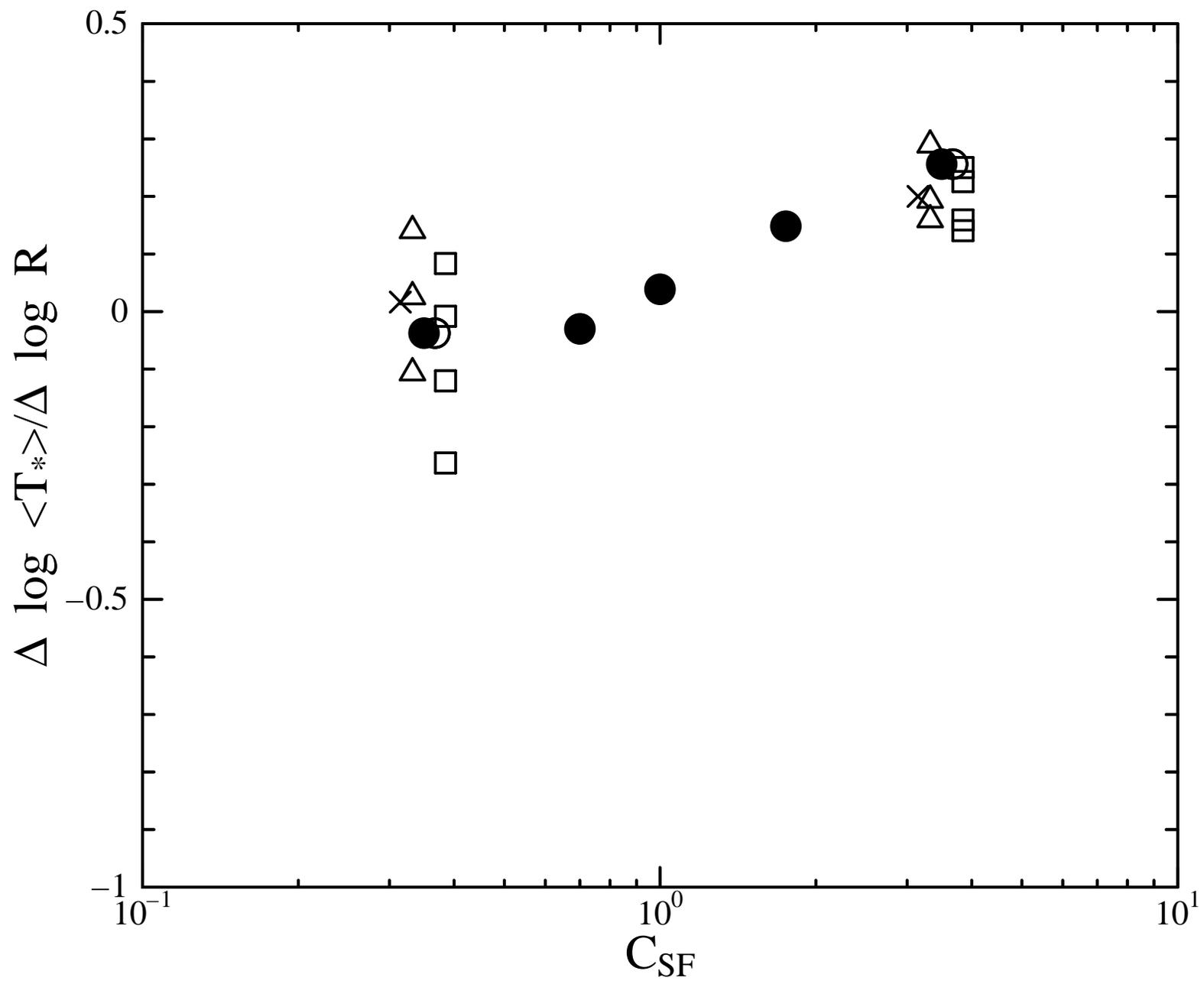
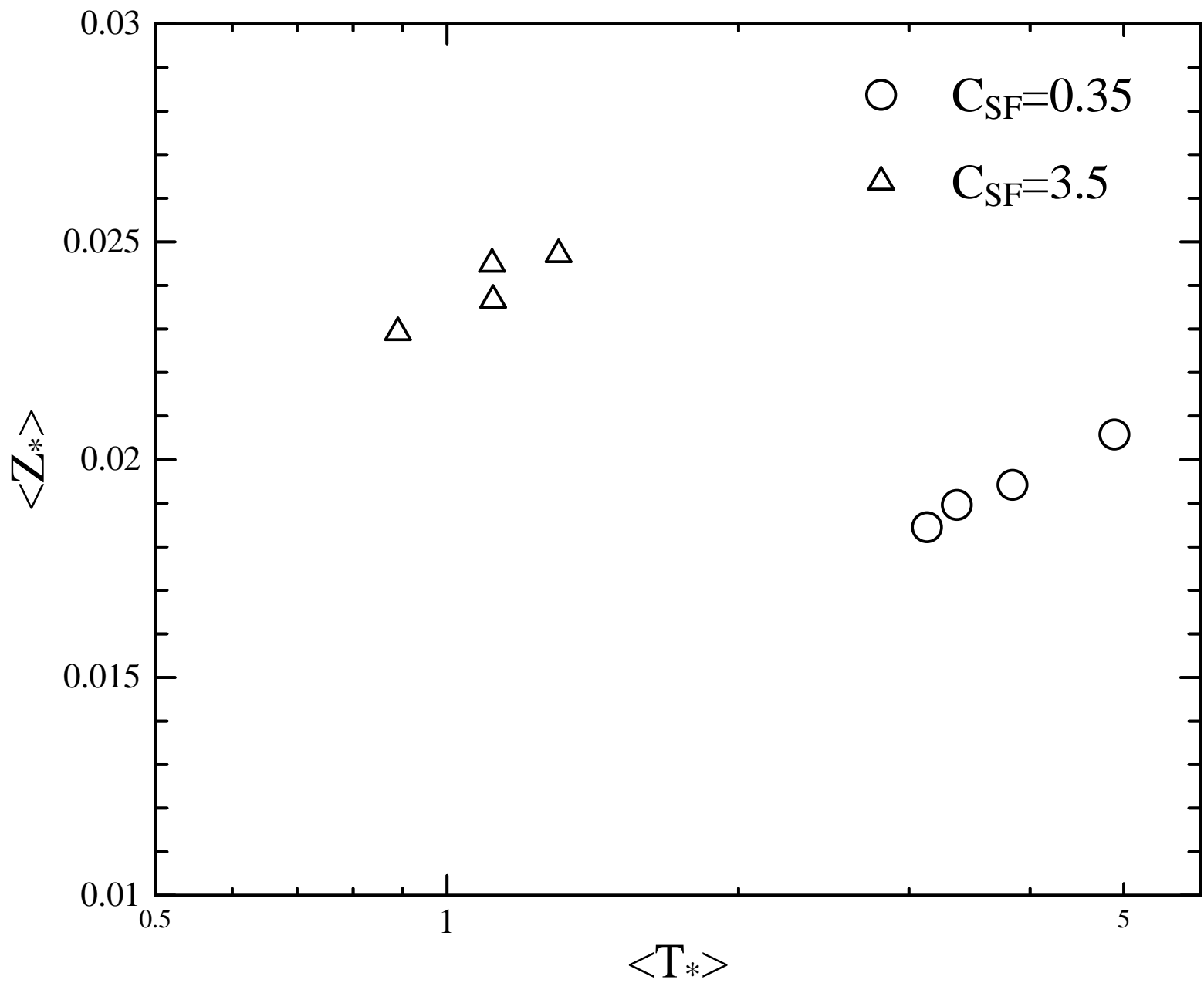
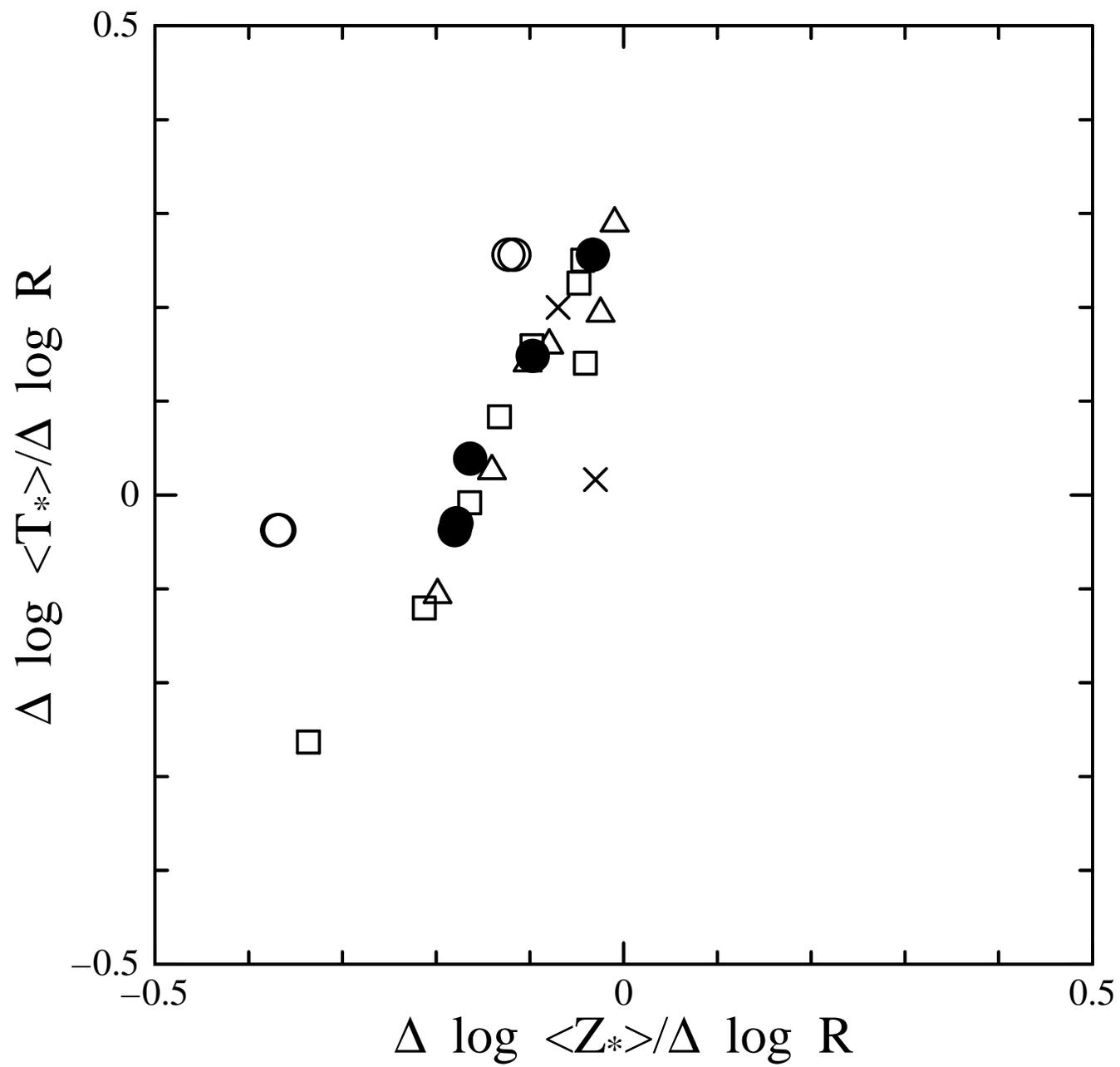
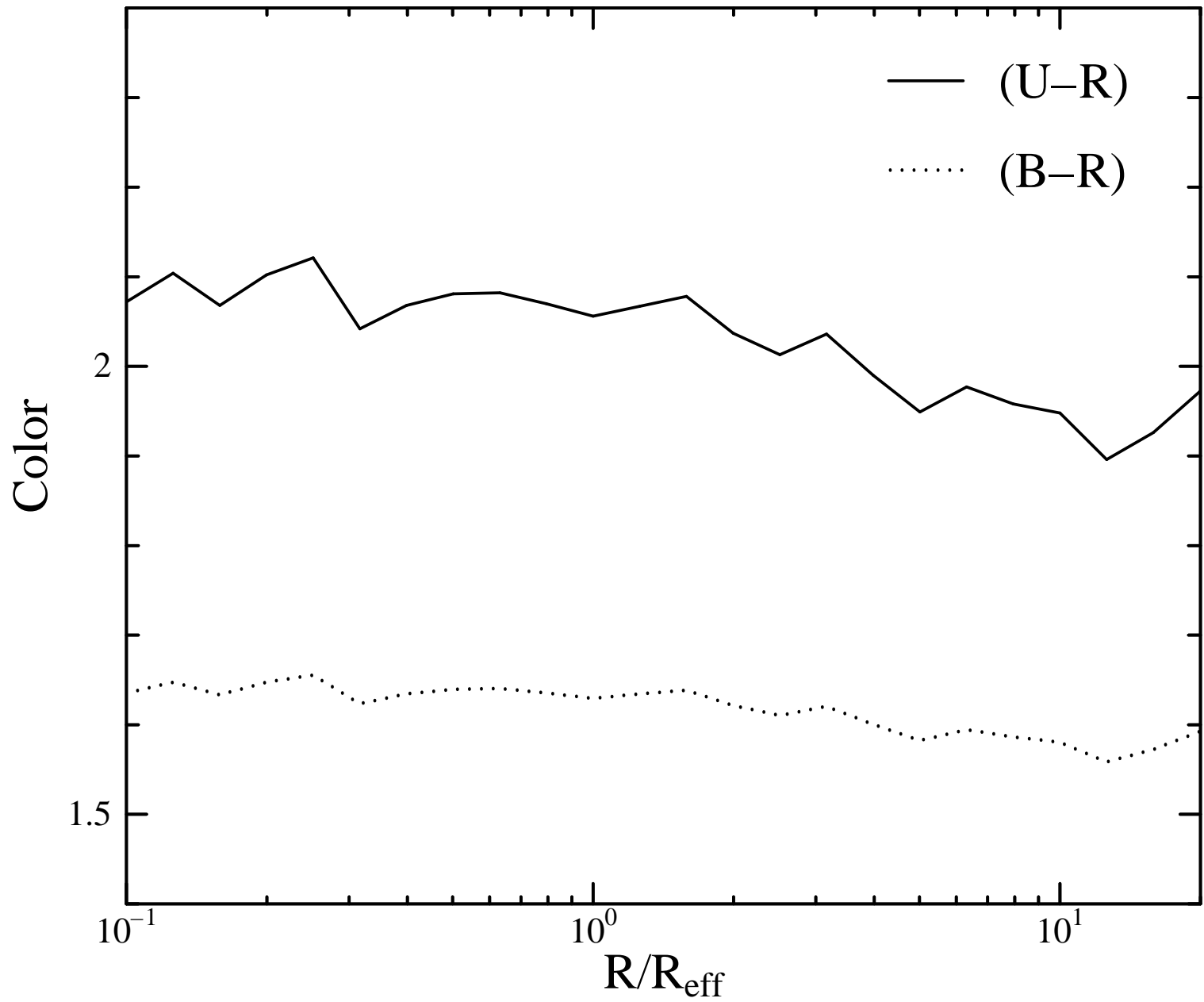
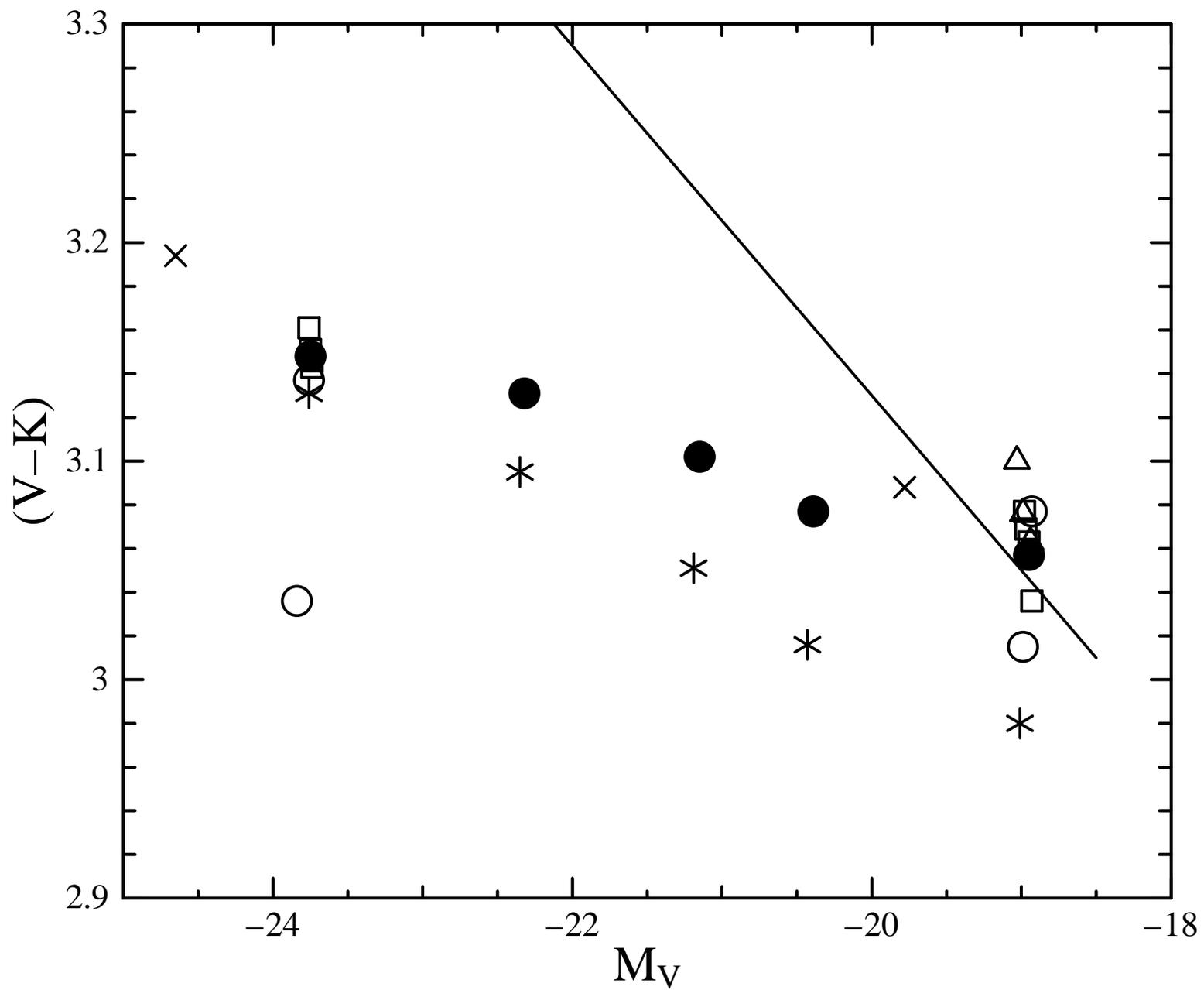
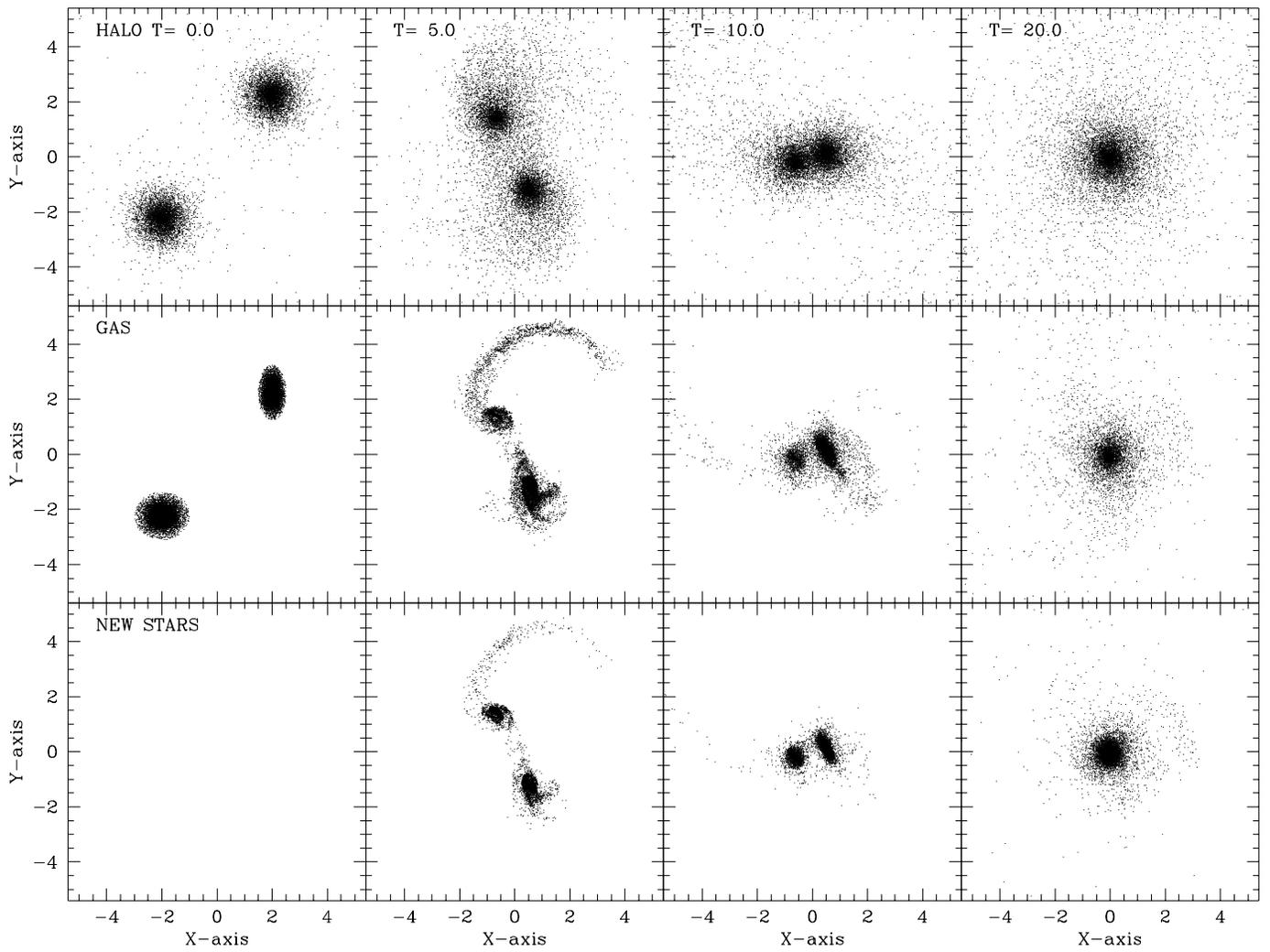
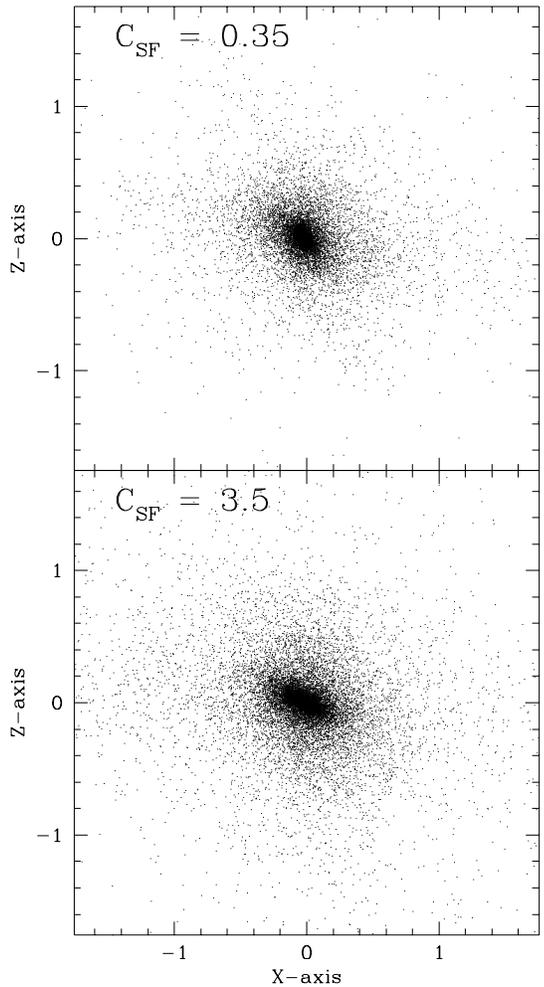
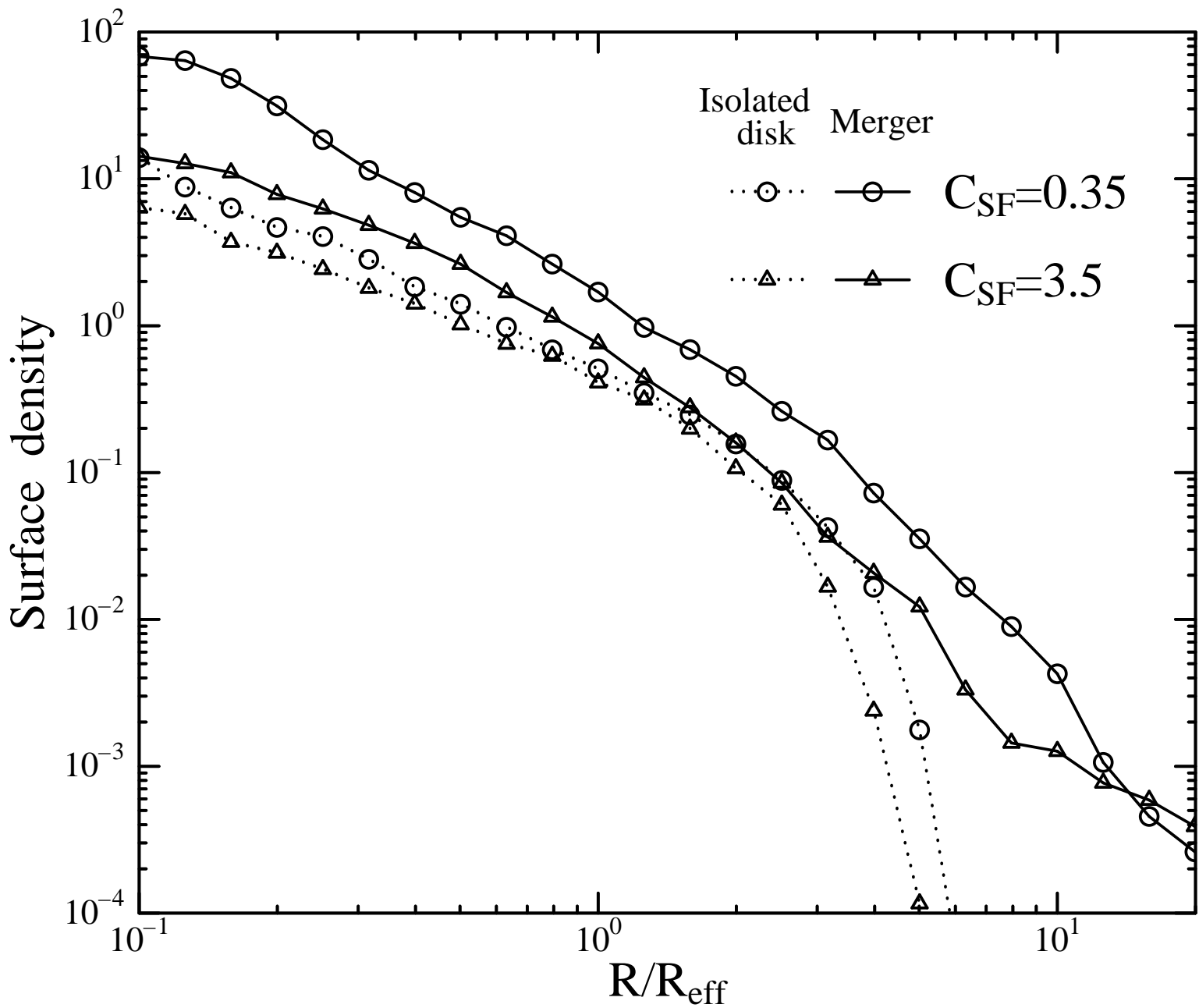
\end{document}